\documentclass[aps,prb,superscriptaddress,reprint,twocolumn]{revtex4}

\usepackage{graphicx}
\usepackage{amsfonts,amsmath,amssymb,amsthm}

\usepackage{graphicx}
\usepackage{amsfonts,amssymb,amsthm}
\usepackage{epstopdf}
\usepackage{upgreek,xspace}
\def\Mu{\ensuremath\upmu\xspace}

\newcommand {\density}{$\times$ 10$^{11}$ cm$^{-2}$}

\newcommand {\gm}{$g^*m^*$/$m_0$}
\newcommand {\Rxx}{$R_\mathrm{xx}$}
\newcommand {\Ryy}{$R_\mathrm{yy}$}
\newcommand {\Rxy}{$R_\mathrm{xy}$}

\newcommand {\EZ}{$g^*\Mu_\mathrm{B}B_\mathrm{t}$}
\newcommand {\EC}{$\hbar eB_\mathrm{p}$/$m^*$}
\newcommand {\EZeeman}{\textit{E}$_\mathrm{Z}$}
\newcommand {\ECyc}{\textit{E}$_\mathrm{cyc}$}
\newcommand {\Bt}{\textit{B}$_\mathrm{t}$}
\newcommand {\Bp}{\textit{B}$_\mathrm{p}$}

\newcommand {\Rs}{$\frac{e^2m^*}{4\pi \hbar^2 \epsilon \sqrt{\pi n}}$}

\newcommand {\Oneup}{$1\uparrow$}

\newcommand {\Zerodown}{$0\downarrow$}
\newcommand {\degree}{$^\circ$}
\newcommand {\ECoul}{\textit{E}$_\mathrm{c}$}
\newcommand{\be}{\begin{equation}}
\newcommand{\ee}{\end{equation}}

\begin{document}

\title{A cascade of phase transitions in an orbitally mixed half-filled Landau level}

\author{J.~Falson}
%\email{j.falson@fkf.mpg.de}
\affiliation{Max-Planck-Institute for Solid State Research, Heisenbergstrasse 1, 70569
Stuttgart, Germany}

\author{D.~Tabrea}
\affiliation{Max-Planck-Institute for Solid State Research, Heisenbergstrasse 1, 70569
Stuttgart, Germany}

\author{D.~Zhang}
\affiliation{State Key Laboratory of Low Dimensional Quantum Physics and Department of Physics, Tsinghua University, Beijing, 100084, China}
\affiliation{Collaborative Innovation Center of Quantum Matter, Beijing, 100084, China }

\author{I.~Sodemann}
\affiliation{Max-Planck-Institute for the Physics of Complex Systems, 01187 Dresden, Germany}

\author{Y.~Kozuka}
\affiliation{Department of Applied Physics and Quantum-Phase
Electronics Center (QPEC), University of Tokyo, Tokyo 113-8656,
Japan}
\affiliation{JST, PRESTO, Kawaguchi, Saitama 332-0012, Japan}

\author{A.~Tsukazaki}
\affiliation{Institute for Materials Research, Tohoku University,
Sendai 980-8577, Japan}

\author{M.~Kawasaki}
\affiliation{Department of Applied Physics and Quantum-Phase
Electronics Center (QPEC), University of Tokyo, Tokyo 113-8656,
Japan} \affiliation{RIKEN Center for Emergent Matter Science
(CEMS), Wako 351-0198, Japan}

\author{K.~von~Klitzing}
\affiliation{Max-Planck-Institute for Solid State Research, Heisenbergstrasse 1, 70569
Stuttgart, Germany}

\author{J.~H.~Smet}
\affiliation{Max-Planck-Institute for Solid State Research, Heisenbergstrasse 1, 70569
Stuttgart, Germany}

%\keywords{Keyword1, Keyword2, Keyword3}

\begin{abstract}
Half-filled Landau levels host an emergent Fermi-liquid which displays an instability towards pairing, culminating in a gapped even-denominator fractional quantum Hall ground state. While this pairing may be probed by tuning the polarization of carriers in competing orbital and spin degrees of freedom, sufficiently high quality platforms offering such tunability remain few. Here we explore the ground states at filling factor $\nu$~=~5/2 in ZnO-based two-dimensional electron systems through a forced intersection of opposing spin branches of Landau levels taking quantum numbers \textit{N}~=~1 and 0. We reveal a cascade of phases with distinct magnetotransport features including a gapped phase polarized in the \textit{N}~=~1 level and a compressible phase in \textit{N}~=~0, along with  an unexpected Fermi-liquid, a second gapped, and a strongly anisotropic nematic-like phase at intermediate polarizations when the levels are near degeneracy. The phase diagram is produced by analyzing the proximity of the intersecting levels and highlights the excellent reproducibility and controllability ZnO offers for exploring exotic fractionalized electronic phases.
\end{abstract}

\flushbottom
\maketitle

\begin{figure*}
\centering
\includegraphics{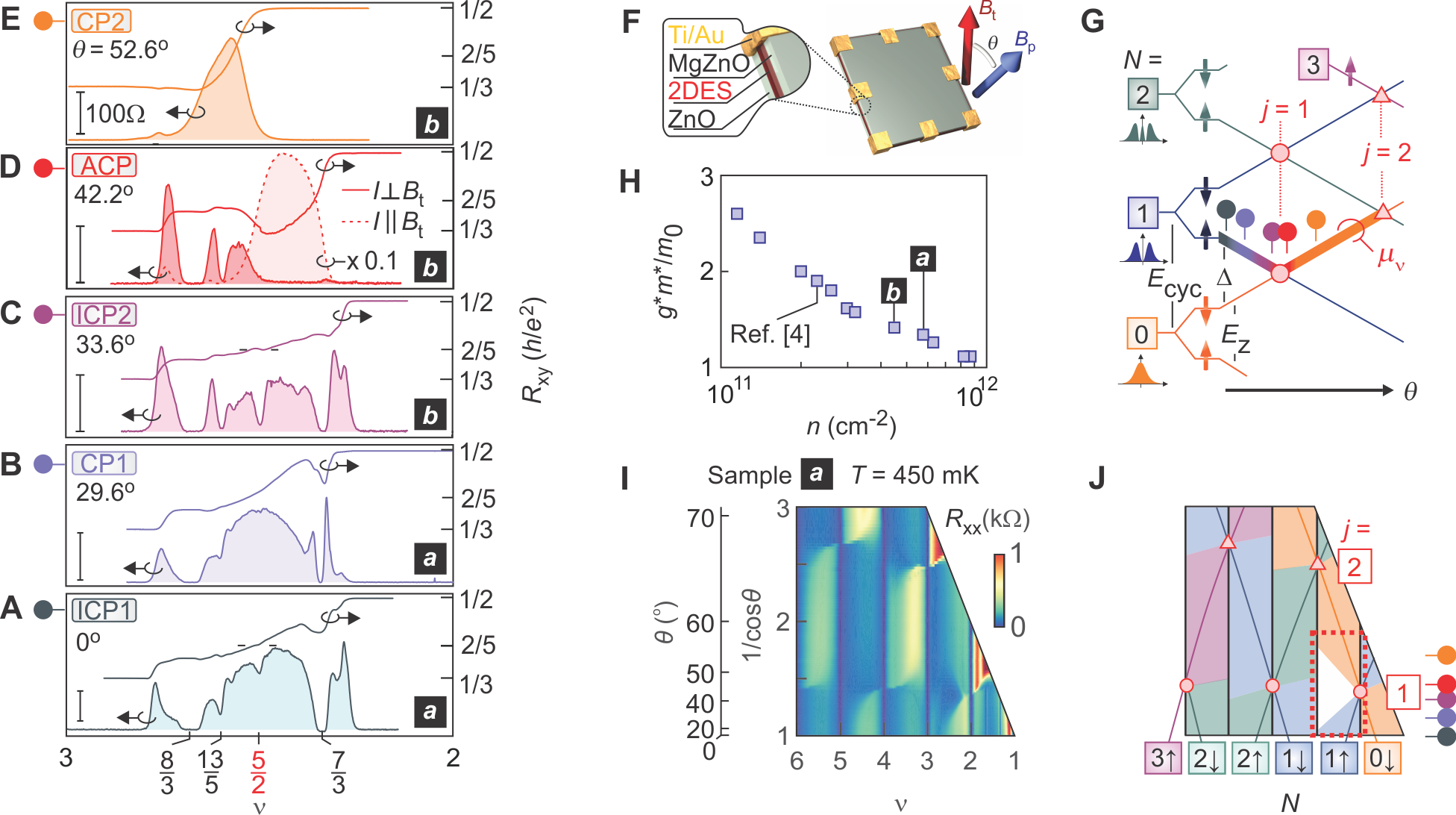}
\caption{\textbf{Overview of correlated phases, samples and the experimental parameter space}. (\textbf{A-E}) Exemplary magnetotransport sweeps of prominent phases explored in this work. The vertical scale bar corresponds to a longitudinal resistance of 100 $\Omega$. (\textbf{F}) The measurement configuration where a MgZnO/ZnO 2DES device is tilted by $\theta$ relative to \Bt. (\textbf{G}) Single particle energy diagram of spin split LL, where $\theta$ is increased and \EZeeman~is selectively enhanced relative to \ECyc~with $\Delta$ their energy difference. Opposing spin levels of adjacent LL cross at the \textit{j}th coincidence position. The chemical potential ($\mu_\nu$) is traced as a bold red line for $\nu$~=~5/2. (\textbf{H}) Spin susceptibility (\gm) as a function of charge density (\textit{n}). (\textbf{I}) \textit{T}~=~450 mK magnetotransport map of sample \textit{a} (\textit{n}~=~5.8 \density) with panel (\textbf{J}) color coding the corresponding orbital and spin character of partially filled levels and \textit{j}~=~1 and 2 coincidence points.}
\label{Fig1}
\end{figure*}

%\cite{willett:1987,rezayi:2000}
%\cite{lilly:1999a,lilly:1999b,pan:1999,samkharadze:2016,liu:2016}
%\cite{jain:2007}

In the presence of strong quantizing perpendicular magnetic fields (\Bp), a large class of correlated phases in two dimensional electron systems (2DES) can be understood as states of weakly interacting composite fermions (CF); an electron bound to an even number of magnetic vortices (fluxes).\cite{jain:2007} Of particular significance is the nature of the ground state that forms in half-filled Landau levels (LL).\cite{willett:1987,ki:2014,falson:2015a,zibrov:2017,li:2017} It is well established that in the half-filled \textit{N}~=~0 LL a correlated metal that can be viewed as a Fermi liquid-like state of CF forms.\cite{halperin:1993} In the \textit{N}~=~1 LL the Moore-Read (MR) state,\cite{read:2000} a gapped phase of CF paired in a $p+ip$-wave Bardeen-–Cooper-–Schrieffer-like fashion, emerges instead. The former is one of the rare examples of a gapless fractionalized phase of matter realized in nature, possibly along with spin liquid candidate materials,\cite{zhou:2017} while the latter is one of the few states with Majorana excitations,\cite{alicea:2012} along with $^3$He films, 1D superconducting wires and spin triplet superconductors like Sr$_2$RuO$_4$. However, among these platforms the MR state is unique in possessing truly non-abelian topological order with fully deconfined point-like non-abelian quasiparticles, although recent studies have advocated for a closely related state arising in the spin liquid candidate $\alpha$-RuCl$_3$ (Ref. \onlinecite{kasahara:2017}). The potential for demonstrating fault-tolerant quantum computation by braiding such quasiparticle excitations has conjured significant interest around their detection and manipulation.\cite{nayak:2008} Insight into the nature of these states may be gained through their study in a regime where LL are well separated, as well as their fate as a pseudospin degree of freedom is tuned and levels are mixed.\cite{falson:2015a,zibrov:2017,li:2017,barkeshli:2016,zaletel:2018} Systems which offer sufficient quality concomitant to experimentally accessible tunability are rare, with bilayer graphene delivering control of the isospin polarization\cite{hunt:2017} of its unique energy spectrum where the \textit{N}~=~0 and 1 LL are nearly degenerate,\cite{novoselov:2006,mccann:2006} and the MgZnO/ZnO-based 2DES through control of the real spin polarization.\cite{falson:2015a} The ability to tune these energy levels also provides a vehicle for exploring intermediate polarization regimes where strong coupling between the pseudospin and real space may result in states with coexisting nematic and ferroelectric-like features.\cite{cote:2010,cote:2011}

Here we explore the ground states of the MgZnO/ZnO-based 2DES at filling factor $\nu$~=~5/2 across a forced crossing of opposing spin branches of \textit{N}~=~1 and 0 levels, where $\nu = \frac{hn}{eB_\mathrm{p}}$, $h$ is the Planck constant, $n$ is the carrier density and $e$ is the elementary charge. At the neighboring integer filling ($\nu$~=~2) it is known that, because of the opposing spin, the transition is a first-order Ising-like spin-flop without any intermediate coherence between the two flavors.\cite{depoortere:2000} It is therefore tempting to speculate that at fractional fillings one would encounter a similarly simple first-order phase transition between the state that is favored for the respective \textit{N}~=~0 and 1 LL. However, what we have uncovered is far more complex: at least five distinct phases emerge as charge is transferred between the \textit{N}~=~1 lower spin and the \textit{N}~=~0 upper spin levels. Figure \ref{Fig1}\textbf{A-E} illustrates these phases using magnetotransport line traces taken on two samples, \textit{a} and \textit{b}. Each trace is recorded when the sample is oriented at a defined tilt angle ($\theta$) relative to the applied magnetic field which acts to increase the spin splitting energy and tune the levels towards and ultimately beyond degeneracy, as will be further discussed below. Panel \textbf{A} displays transport when the chemical potential ($\mu_\nu$) lies in \textit{N}~=~1 and reveals an incompressible phase at $\nu$~=~5/2, which we denote as ICP1. Upon increasing $\theta$, this state evolves into a compressible phase (panel \textbf{B}), noted as CP1. Then, for a narrow but finite range of $\theta$ a second incompressible phase at $\nu$~=~5/2 emerges (ICP2, panel \textbf{C} taken on sample \textit{b}). This state is supplanted by a strongly anisotropic yet compressible phase (ACP), as per panel \textbf{D} which displays the longitudinal resistance taken when the current (\textit{I}) is sent in two orthogonal crystal directions. Finally, upon full polarization in \textit{N}~=~0 at high $\theta$ the partial filling is rendered compressible (CP2, panel \textbf{E}). In this manuscript we convey the most poignant experimental findings using two samples but stress that the phase transition cascade is highly reproducible, as summarized using five samples in Fig. \ref{Fig4}. A range of temperatures are used to illustrate certain characteristics: low temperature (\textit{T}~$=$~20 mK) for ground states and elevated temperatures for charge density wave physics (\textit{T}~=~90 mK), hysteretic transport (\textit{T}~=~200 mK) and comprehensive mapping of level crossings (\textit{T}~=~450 mK). Finally, we conclude with a discussion on the possible ground states realized throughout the cascade. 

For a certain sample we tune between these phases by tilting the MgZnO/ZnO-based 2DES within a magnetic field (Fig. \ref{Fig1}\textbf{F}). This is a commonly employed technique in 2D systems to selectively enhance the Zeeman energy (\EZeeman~=~\EZ, where \textit{g}$^*$ is the isotropic effective \textit{g}-factor and $\Mu_\mathrm{B}$ is the Bohr magneton) relative to the cyclotron gap (\ECyc~=~\EC, $\hbar$ is the reduced Planck constant and $m^*$ is the effective mass) as the former depends on the total field (\Bt) and the latter only on the perpendicular component. Their ratio scales as

\begin{equation}
\frac{E_\mathrm{Z}}{E_\mathrm{cyc}} = \frac{g^*m^*}{2m_0\mathrm{cos}\theta} = j.
\label{Eq.Coincidence}
\end{equation} 

\noindent
Here, \gm~is the spin susceptibility of carriers and governs the energy spacing ($\Delta$) of levels at $\theta$~=~0$^\circ$, as illustrated in Fig.\ref{Fig1}\textbf{G} using the single particle energy ladder of spin split LL. Following from Eq.\ref{Eq.Coincidence}, \EZeeman/\ECyc~may reach an integer value upon increasing $\theta$ which will result in a level crossing, where \textit{j} is the difference in orbital index of the coinciding levels. Such a crossing will result in a discrete change in orbital character and spin projection of carriers at $\mu_\nu$. 

In addition to tuning $\theta$, appropriate sample selection is essential for accessing the desired experimental parameter space for probing these crossings. This is due to the renormalization of \gm~in an interacting 2DES as \textit{n} is reduced and the magnitude of the Coulomb energy is amplified relative to the zero-field kinetic energy of carriers (Ref. \onlinecite{gokmen:2010}). In contrast to a density-independent \gm~predicted for non-interacting electrons, we observe a nearly three-fold increase of \gm~in the ZnO-based heterostructures when reducing \textit{n} from 10$^{12}$ to 10$^{11}$ cm$^{-2}$ as shown in Fig.\ref{Fig1}\textbf{H}. This can be viewed as the Landau parameters drifting towards the critical value for the appearance of a Stoner-type instability in the dilute limit. Therefore, setting \textit{n} will determine an initial $\Delta$ of the sample which then may be continuously modified by tuning $\theta$.

The single particle level diagram in Fig.\ref{Fig1}\textbf{G} is efficient in capturing the general constellation of level crossings in the experiment, as shown in Fig. \ref{Fig1}\textbf{I} for sample \textit{a}. The data is gathered by sweeping \Bt~at a set $\theta$, which is later used to calculate \Bp~and $\nu$. Quantum Hall features are seen as dark blue vertical lines. Here, we use an elevated temperature of \textit{T}~$=$~450 mK to suppress correlated ground states while accentuating the distinct chequer-board pattern which emerges. Previous studies\cite{maryenko:2014,falson:2015a} have shown this to be associated with the spin of electrons at $\mu_\nu$, with high (low) resistance corresponding to majority (minority) spin ($\sigma = \uparrow,\downarrow$) carriers. The partial filling factor always ends up high resistance upon full spin polarization, as can be seen at high $\theta$ for $\nu<$3. A comparison with the single particle energy ladder allows us to assign ($N,\sigma$) quantum numbers to the partially filled levels at $\mu_\nu$ as shown in Fig.\ref{Fig1}\textbf{J}. 

\begin{figure}
\centering
\includegraphics{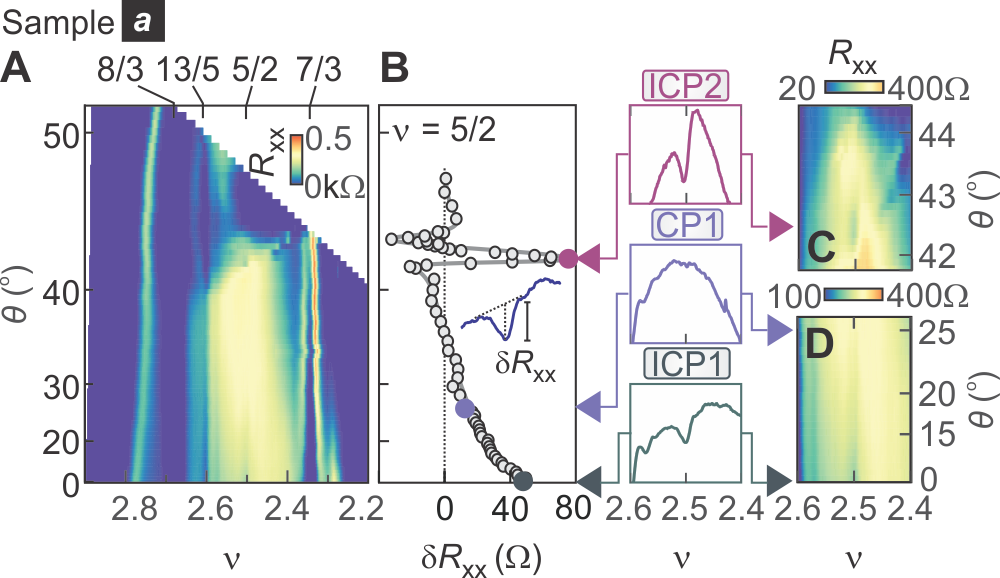}
\caption{\textbf{Ground states of sample \textit{a}.} (\textbf{A}) \textit{T}~$=$~20 mK magnetotransport map. (\textbf{B}) Minimum of resistance ($\delta$\Rxx)~at $\nu$~=~5/2 relative to surrounding fillings as a function of $\theta$. A positive value indicates a minimum at $\nu$~=~5/2, as shown in the inset data trace. Close up mapping of the (\textbf{C}) ICP2 and (\textbf{D}) ICP1 and CP1 phases with boxes displaying magnetotransport traces at the angle indicated by colored arrows. The box height corresponds to $\Delta$200$\Omega$.}
\label{Fig2}
\end{figure}

We now focus on the ground states for 3~$>\nu>$~2 filling across the \textit{j}~=~1 coincidence between the \Oneup~and \Zerodown~levels. This region is framed in the dotted red box in Fig.~\ref{Fig1}\textbf{J}. We present data from two samples for the following reasons: sample \textit{a} (\textit{n}=5.8~\density, \gm~$\approx$~1.35) has a large $\Delta$ which allows us to explore a larger portion of transport in the \textit{N}~=~1$\uparrow$ level but limits the maximum accessible $\theta$ at $\nu$~=~5/2 due to the experimentally available \textit{B}-field. Sample \textit{b} (\textit{n}=4.4~\density, \gm~$\approx$~1.48) however permits us to observe transport well beyond the first coincidence position, due to its lower \textit{n}, and in greater detail thanks to its higher mobility. Reducing \textit{n} in sample \textit{b} however acts to suppress $\Delta$ which limits the extent transport in the $N=1\uparrow$~level may be observed. 

\begin{figure}
\centering
\includegraphics{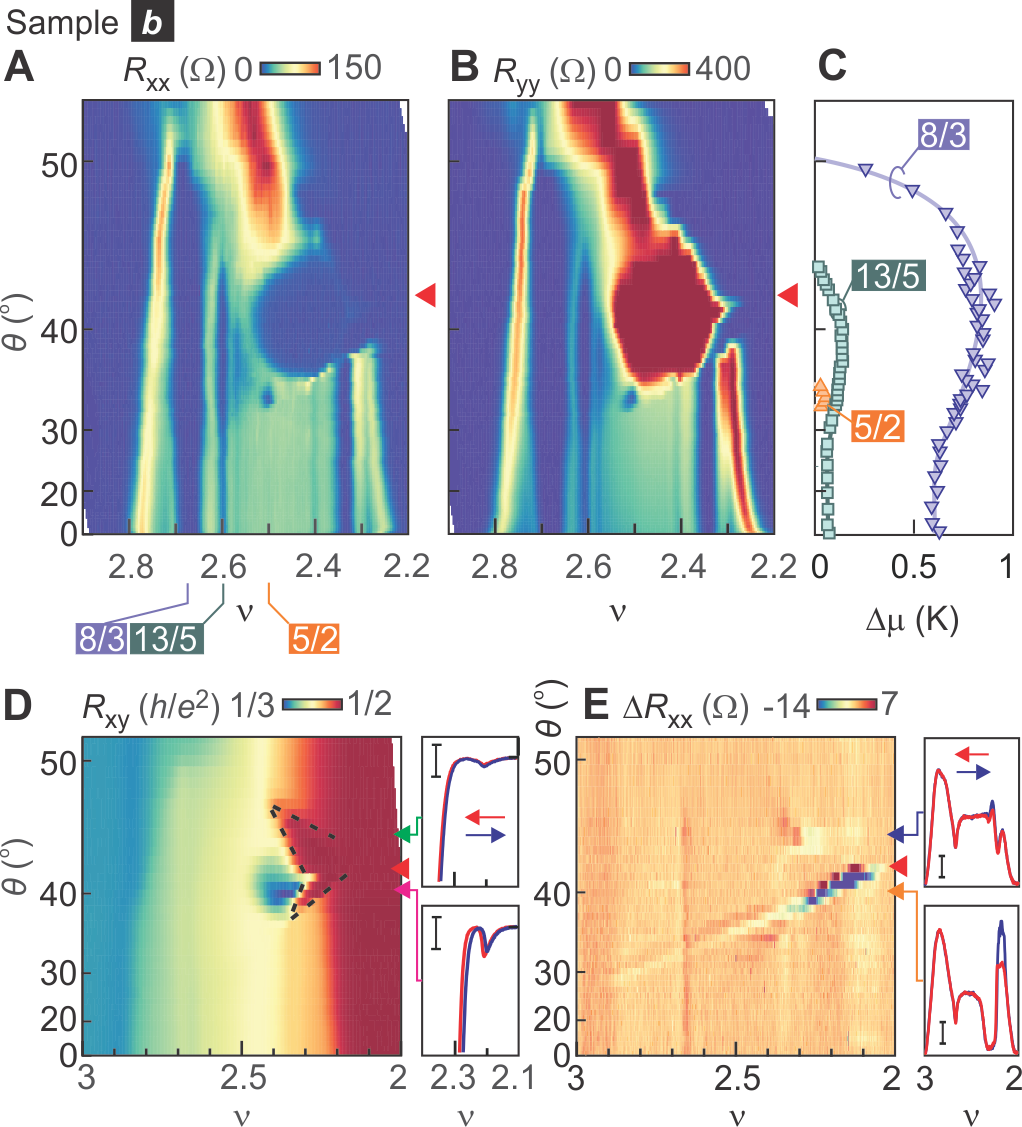}
\caption{\textbf{Ground states and hysteresis of sample \textit{b}.} (\textbf{A}) and (\textbf{B}) Magnetotransport maps (\textit{T}~$=$~20mK) for two orthogonal crystal directions, with \Rxx~associated with \textit{I}$\perp$\Bt~and \Ryy~with \textit{I}$\parallel$\Bt. (\textbf{C}) Activation energy $\Delta_\mu$ of prominent FQH states. (\textbf{D}) Isotropic reentrant integer quantum Hall features at \textit{T}~=~90 mK: mapping of \Rxy~with reentrant integer quantum Hall features of $\nu$ = 2 framed by dotted lines. The sub-panels identify their hysteretic behavior with the scale bar equalling 0.1[\textit{h}/\textit{e}$^2$]. (\textbf{E}) Hysteresis at \textit{T}~=~200 mK: mapping of $\Delta$\Rxx. Individual traces with \textit{B}-sweep direction dependent transport are shown for discrete angles in the sub-panels. The scale bar corresponds to an \Rxx~of 10$\Omega$. Red triangles indicate the \textit{j}~=~1$|_{\nu=2}$ coincidence position.}
\label{Fig3}
\end{figure}

\begin{figure}[h]
\centering
\includegraphics{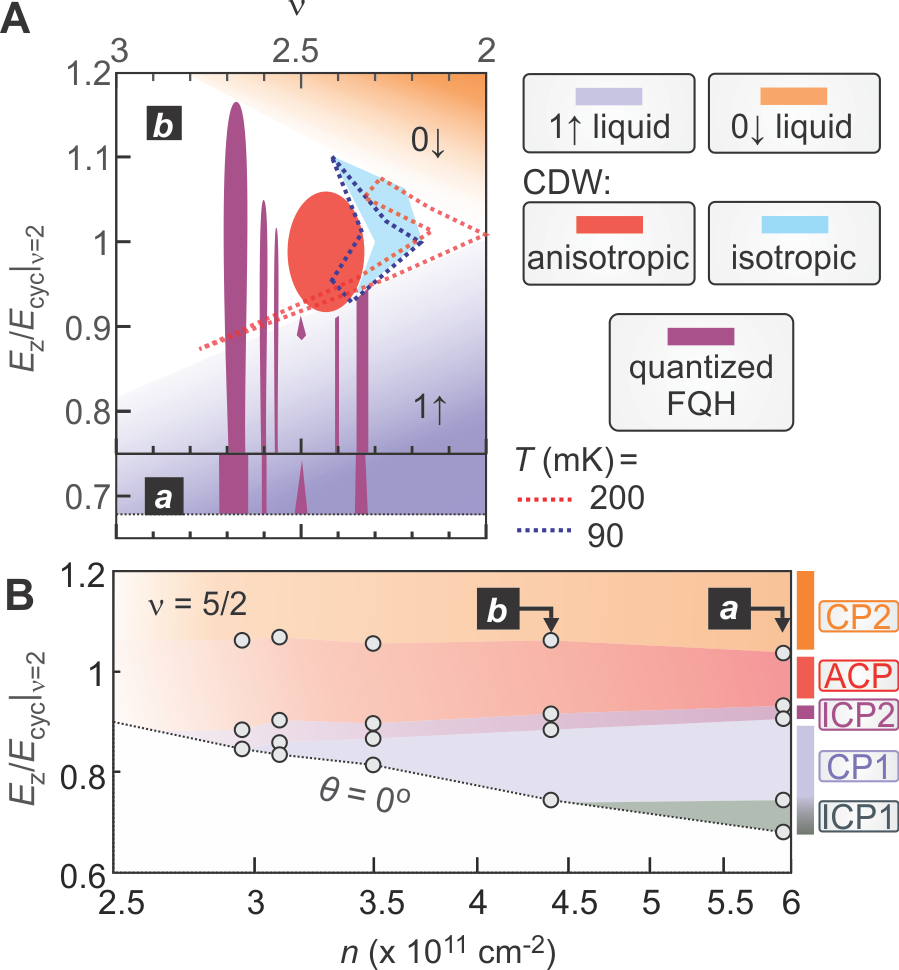}
\caption{\textbf{Phase diagram of ground states.} (\textbf{A}) Map of the ground states in the ($\nu$--\EZeeman/\ECyc)-plane through the transitions from \textit{N} = 1$\uparrow$~(light purple) to 0$\downarrow$~(orange) orbital character. The range of FQH features are shown as purple, while the anisotropic CDW-like phase is red and isotropic in light blue. The range of hysteretic features identified in transport maps are shown as dotted lines for \textit{T}~=~200 (red) and 90~mK (blue). (\textbf{B}) A summary of the transitions at $\nu$~=~5/2 in terms of \EZeeman/\ECyc$|_{\nu=2}$ as a function of \textit{n}, as per Eq. \ref{Eq.Coincidence}.}
\label{Fig4}
\end{figure}

Figure \ref{Fig2}\textbf{A} maps the magnetotransport of Sample \textit{a} in the ($\nu$--$\theta$)-plane at base temperature. Panel \textbf{B} tracks the minimum of \Rxx~at $\nu$~=~5/2 relative to the surrounding fillings of this map. A positive value of $\delta$\Rxx~corresponds to a minimum at $\nu$=5/2. The ICP1 phase (panel \textbf{D}) is resolved at low $\theta$ and gradually evolves into the CP1 phase by $\theta\approx$25$^\circ$. The ICP2 phase appears at $\theta\approx 42.5^\circ$, as per panel \textbf{C}, and is seen as a spike in $\delta$\Rxx~in panel \textbf{B}. At slightly higher $\theta$ a low-resistance region is resolved and is associated with the ACP which will be discussed further using Fig. \ref{Fig3}.

Figure \ref{Fig3} presents magnetotransport mapping of sample \textit{b} whose intrinsically smaller $\Delta$ renders the CP1 phase stable at $\theta$~=~0$^\circ$. Starting from this phase, the same sequence of states at higher $\theta$ is seen. The anisotropic transport at larger $\theta$ is demonstrated by contrasting the data of Fig.~\ref{Fig3}\textbf{A} and \textbf{B} where the resistance is taken along two orthogonal crystal directions. The ACP is the ground state for $35^\circ \lessapprox \theta \lessapprox 45^\circ$ and $2.3 \lessapprox \nu \lessapprox 2.55$. The ACP phase exhibits easy-axis transport features when \textit{I}$\perp$\Bt, which evidently acts as a symmetry breaking field as the anisotropic features will re-orient by 90$^\circ$ if the projection of \Bt~is also rotated. This range of $\theta$ also exhibits alternative flavors of charge density wave (CDW)-like physics: reentrance of \Rxy~into the integer quantum Hall condition at $\nu$~=~2 is seen fanning to higher and lower $\theta$, as marked by the dotted lines in Fig.~\ref{Fig3}\textbf{D} at \textit{T}~=~90 mK. The higher mobility of sample \textit{b} allows us to plot the activation gaps ($\Delta_\mu$) of prominent fractional quantum Hall (FQH) features as a function of $\theta$ (panel \textbf{C}). The ICP2 phase is realized for a particularly narrow range of $\theta$, beyond which the odd-denominator states of 8/3 and 13/5 display an increase in stability until the system enters the CP2 phase and carriers at $\mu_\nu$ are polarized in \textit{N}~=~0$\downarrow$.

The crossing at $\nu$~=~2 is an Ising transition between two oppositely polarized ferromagnetic states and may be resolved in transport due to the emergence of hysteresis associated with domain formation.\cite{depoortere:2000} The charge transfer dynamics at fractional filling factors is more elusive as hysteresis is either weak or absent. We can gain a glimpse of this by inspecting the hysteretic component of transport at an elevated temperature of \textit{T}~=~200 mK as shown in Fig.~\ref{Fig3}\textbf{E}. This data is constructed by taking two transport maps, one comprised of only down-sweep and the other only up-sweep \textit{B}-field data, with the difference ($\Delta$\Rxx) being displayed. Strong hysteretic features are observed close to $\nu$=~2 and is evidence for a first-order transition associated with the exchange corrected single-particle level crossing at \textit{j}~=~1(red arrow, see SI).\cite{giuliani:2005} These dramatically weaken at fractional fillings of $\nu > 2.3$. We also note the weaker features in the sub-panels of \textbf{D} and \textbf{E}, which emerge in the vicinity of reentrant features at both higher and lower $\theta$ relative to the \textit{j}~=~1 crossing position.

Figure \ref{Fig4} summarizes the phase diagram of competing ground states. Panel \textbf{A} is a composite representation of the extent of the ground states in the ($\nu$--\EZeeman/\ECyc)-plane for sample \textit{a} and \textit{b}, including FQH (dark purple) features, \Oneup~(light purple) and \Zerodown~(orange) polarized liquids and anisotropic (red) and isotropic (blue) CDW-like phases. For the sake of constructing this diagram, we determine \EZeeman/\ECyc~using the \gm~quantified at \textit{j}~=~1$|_{\nu = 2}$ and using Eq. \ref{Eq.Coincidence}. This representation allows us to define a single-particle energy scale for the sake of discussion and comparison (see SI). Extrapolating the hysteretic features at a set temperature observed near $\nu$~=~2 at \textit{T}~=~200 mK (red) (Fig.~\ref{Fig4}\textbf{A}) and 90 (blue) mK (see SI) permits us to illustrate a range in which the 2DES is depolarized and has mixed \textit{N}~=~0 and 1 character. Interestingly, the $\nu$~=~8/3 and 13/5 states do not close their charge gap across the lower bound of this mixed regime (Fig.\ref{Fig3}\textbf{C}). Moreover, the ICP2 and ACP phases emerge only within these bounds. These observations support our hypothesis that the charge transfer at fractional fillings is unique and gradually occurs over a large range of $\theta$ rather than at a single discrete angle. Panel \textbf{B} highlights the reproducibility of the phase diagram by plotting the cascade as a function of \textit{n}; the \EZeeman/\ECyc~ratio emerges as the determining factor of the stability of ground states for all samples. It also conveys why these phases evaded detection in previous works:\cite{falson:2015a} the sample investigated (\textit{n}~=~2.3\density, \gm~=~1.9) put the majority of phases out of experimental reach due to an intrinsic \EZeeman/\ECyc~ratio of approximately 0.95.

Henceforth, we discuss candidate states for the observed phases by employing the CF picture at $\nu$~=~5/2. We will construct states by adding holes to the neighboring integer quantum Hall state at $\nu$~=~3, and thus we write $\nu = 3-(\nu_{h0}+\nu_{h1})$, with $\nu_{h0}+\nu_{h1}$=1/2 the partial hole fillings of the \textit{N}~=~0($\downarrow$)/1($\uparrow$) components. When $\Delta$ is large (\textit{n} is large), the ICP1 phase is likely the MR state\cite{moore:1991} for $\nu_{h0}=0$ and $\nu_{h1}=1/2$. An interesting possibility as the crossing is approached is a mixed state known as the $Z_2$ exciton metal with the coexistence of pairing of the \textit{N}~=~1 component and a composite Fermi liquid in the \textit{N}~=~0 component.\cite{barkeshli:2016,zibrov:2017,zaletel:2018} This state would have $R_\mathrm{xx}\rightarrow$0 and $R_\mathrm{xy}[h/e^2]=2/5$ as \textit{T}$\rightarrow$0 (see SI), and hence not ruled out by our observations although measurement of the spin polarization will be needed to confirm its presence.

The CP2 phase naturally corresponds to $\nu_{h0}=1/2,~\nu_{h1}=0$. Surprisingly, odd-denominator FQH physics is entirely absent in this regime, despite the \textit{N}~=~0 orbital character. We attribute this to the reduced screening of the polarized state that stems from an increased Pauli blocking of virtual transitions. This reduced screening enhances the effective disorder potential, degrading the quality of the odd-denominator FQH features and increasing the width of the $\nu$~=~2 plateau. 

The physics of CP1 is subtle as it emerges in the vicinity of the level crossing of \textit{N}~=~0$\downarrow$ and 1$\uparrow$ as $\Delta$ is made small. Accordingly, we speculate that this phase corresponds to a state with two Fermi surfaces ($(\nu_{h0},\nu_{h1})\ne$0). States with two composite Fermi surfaces are known to be energetically favored in the case of spin-full \textit{N}~=~0 LL with small Zeeman splitting.\cite{kukushkin:1999,tiemann:2012} Therefore it is not unreasonable that partially polarized two component Fermi sea states might arise in our case of a spinful two component orbitally mixed system (see Refs. \onlinecite{zibrov:2017},\onlinecite{barkeshli:2016},\onlinecite{zaletel:2018} and a further discussion in the supplementary information).

The nature of the ICP2 phase constitutes another intriguing puzzle. One possibility is a paired MR state, which has been found to exhibit enhanced stability near level crossings of subbands in GaAs.\cite{liu:2011} Another alternative is that ICP2 is a paired state distinct from the MR state. A natural candidate would be the Halperin-331 state, which is two-component and could therefore be favored near the level crossing.\cite{read:2000}

Finally, we discuss the CDW-like physics which emerges between ICP2 and CP2. It is important to emphasize that the ACP phase is not restricted to 5/2 but pervades a large portion of the filling fraction range. A natural candidate is the stripe phase seen in higher LL (\textit{N} $\geq$ 2) in GaAs.\cite{fogler:1997,lilly:1999a} There, the FQH effect is expelled as CDW physics becomes energetically favorable. Similar features can be induced in \textit{N}~=~1 as the MR state is in close competition with a stripe phase.\cite{lilly:1999b,pan:1999,rezayi:2000,samkharadze:2016} What makes the ACP state in this work unexpected is its stabilization near the level crossing as \textit{N}~=~0 character is increased, as evidenced by the increase in activation energy of odd-denominator states at $\nu$~=~8/3 and 13/5 for the same range of $\theta$ (Fig. \ref{Fig3}\textbf{C}). The anisotropy is also notably askew towards low $\nu$ where isotropic reentrant features additionally emerge. An interesting alternative scenario to conventional stripes, is that ACP corresponds to a coherent state between \textit{N}~=~0$\downarrow$ and 1$\uparrow$ levels. As these orbitals carry different angular momentum, a coherent superposition would break inversion and rotation symmetries. This would endow the state with nematic and ferroelectric characteristics, explaining its anisotropic nature. In addition, the opposite spin would render the state with a finite magnetization in the plane orthogonal to the spin polarization axis dictated by \EZeeman. While proposals for related ferroelectric states in bilayer graphene have been made at integer fillings,\cite{cote:2010} we are not aware of studies addressing their energetic feasibility for fractionally filled levels. We hope that our results motivate future numerical and experimental studies of such possibilities.

In summary, the exotic sequence of phases observed at the crossing of the \textit{N}~=~0 and 1 LL is consistent with a gradual and complex depolarization sequence of the orbital character, in contrast to the expectations of a simple spin-flop first-order transition. In addition to LL-polarized states, the unexpected observation of incompressible and anisotropic phases in the orbitally mixed regime opens important questions concerning their nature and the possibility of novel inter-layer coherence arising at fractional filling factors. \\

\textbf{Methods} The Mg$_x$Zn$_{1-x}$O/ZnO heterostructures were grown using ozone molecular beam epitaxy.\cite{falson:2018} Sample \textit{a} (\textit{n}~=~5.8 \density, \textit{x} $\approx$ 0.04, \gm~=~1.35) and \textit{b} (\textit{n}~=~4.4 \density, \textit{x} $\approx$ 0.03, \gm~=~1.48) have electron mobilities of 190,000 cm$^2$/Vs and 350,000 cm$^2$/Vs. Measurements were performed in a top-loading-into-the-mixture ($T_\textrm{base} \approx $ 20 mK) dilution refrigerator equipped with a rotation stage. Low frequency AC lock in techniques were used to gather resistance data. \\

\iffalse
\textbf{Data availability} Data is available upon reasonable request from the authors.\\

\textbf{Author contributions} J.F, D.T and D.Z gathered the transport data. J.F grew the samples with assistance from Y.K, A.T and M.K. J.F and I.S wrote the manuscript following discussion and input from all authors.\\

\textbf{Competing interests} The authors declare no competing financial interests.

\fi
\textbf{Acknowledgments} We appreciate discussions with M. Barkeshli, B. Feldman, Yang~Liu and M. Zudov. We acknowledge the financial support of JST CREST Grant Number JPMJCR16F1, Japan. J.F. acknowledges the Max~Planck$-$University of British Columbia$-$University of Tokyo Center for Quantum Materials and the Deutsche Forschungsgemeinschaft (FA 1392/2-1). Y.K. acknowledges JST, PRESTO Grant Number JPMJPR1763, Japan.\\

\clearpage

%\onecolumngrid
\section*{SUPPLEMENTARY MATERIALS}

\section*{Further discussion on the ground states}\label{furtherdiscussion}

\textit{General considerations}------To understand the physics near the crossing of $N = 0 \downarrow$ and $N=1\uparrow$, we consider an ideal limit with only these two levels and any other Landau levels neglected. It is convenient to define a filling factor of these two components, $\tilde{\nu}$, which is related to the total filling factor simply by $\nu=1+\tilde{\nu}$, and that ranges from $0 \leq \tilde{\nu} \leq 2$. The region that we have focused on experimentally is $1 \leq \tilde{\nu} \leq 2$, and in particular $\nu = 5/2$ corresponds to $\tilde{\nu}=3/2$.

We call $\Delta$ the single particle splitting between these levels. This splitting is understood to include exchange corrections from the fully occupied Landau levels. In the ideal limit in which Landau level mixing can be neglected, we can project the interacting Hamiltonian into the two level system. The single particle energy of the partially filled $N=1\uparrow$ will acquire a correction due to its exchange interactions with the fully occupied $N=0 \uparrow$ level, however the $N=0 \downarrow$ single particle energies receive no exchange corrections. Therefore in the limit of no Landau level mixing the effective single particle splitting between $N=1\uparrow$ and $N=0 \downarrow$ is:

\be
\Delta=E_1-E_0=\hbar \omega_c+\frac{e^2}{\epsilon l} \epsilon_e(1)-E_Z.
\ee

\noindent Where $\epsilon_e(1)\approx - 0.63$ (Ref. \onlinecite{giuliani:2005}). The Hamiltonian includes also Coulomb interactions projected into these two Landau levels in addition to this single particle splitting. Consider an anti-unitary particle-hole conjugation, $C$, implemented as follows:

\be
C c^\dagger_{nm} C^{-1} = c_{nm}, \ C i C^{-1}=-i.
\ee

\noindent This operation maps states with fillings $\tilde{\nu}\rightarrow 2-\tilde{\nu}$ and reverses the splitting $\Delta\rightarrow -\Delta$ while leaving the interacting part of the Hamiltonian invariant. We are focusing only in the region $1 \leq \tilde{\nu} \leq 2$, and hence we cannot use this operation to relate different fillings within this range. However, the existence of this mapping makes convenient to often describe our states as states of holes, with a hole filling defined as $\nu_h=2-\tilde{\nu}$.

%Because of this operation one can view states of holes in similar way as those of electrons within this two-component space.

The total number of particles in each component is conserved exactly due to conservation of spin (because of the smallness of the spin orbit coupling effects in ZnO), which is an important difference with respect to the case of bilayer graphene. The ground state at any value of $\Delta$ has a well defined polarization given by:

\be
p\equiv\frac{\tilde{\nu}_0-\tilde{\nu}_1}{\tilde{\nu}_1+\tilde{\nu}_0}
\ee

One key question is the dependence of $p$ as a function of $\Delta$. The answer to this question depends on the filling factor. At the level of Hartree-Fock theory at total filling $\tilde{\nu}=\tilde{\nu}_1+\tilde{\nu}_0=1$, one obtains a first-order Ising transition with no intermediate coherence, namely, $p=-1$ for $\Delta<0$ and $p=+1$ for $\Delta>0$. However, whether a sudden or gradual reversal of polarization occurs at fractional fillings away from $\tilde{\nu}=1$ is an unresolved theoretical question.

% for $\Delta<\Delta_c^{\tilde{\nu}=2}$ and $p=+1$ for $\Delta>\Delta_c^{\tilde{\nu}=2}$, where $\Delta_c^{\tilde{\nu}=2}$ is a parameter characterizing the difference of exchange energy corrections to the completely filled $N=1$ and $N=0$ levels. However, whether a sudden flip of polarization or a gradual depolarization is present at fractional fillings away from $\tilde{\nu}=1$ remains an unsettled theoretical question.

\begin{figure}[h]
\centering
\includegraphics{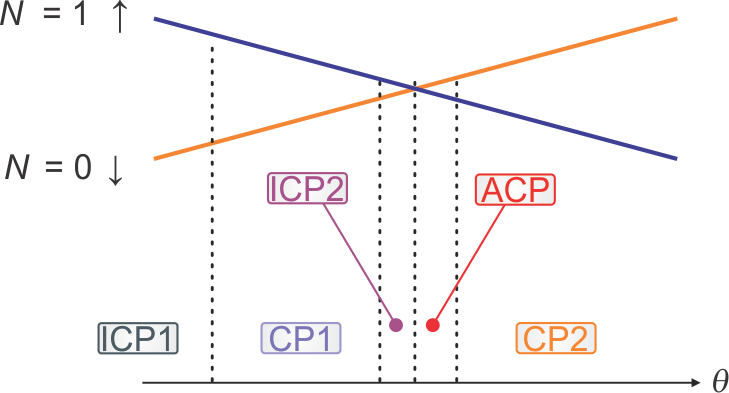}
\caption{Observed sequence of states at filling fraction $\tilde{\nu}=3/2$, which corresponds to a total filing $\nu=5/2$.}
\label{0-1}
\end{figure}

Our experiment has found at $\tilde{\nu}=3/2$ a non-trivial sequence of phase transitions as a function of $\Delta$ which is experimentally modified by changing $\theta$, as summarized in Fig.~\ref{0-1}. We have no direct way of measuring $p$, but we believe such non-trivial sequence is compatible with a partial depolarization taking place in stages as the two levels cross, as we argue in more detail below. Observations of a gradual depolarization in the related system of bilayer graphene were reported in Ref.~\onlinecite{zibrov:2017}.

\textit{Candidate phases}--- We begin by labeling the states we have experimentally encountered at total filling $\tilde{\nu}=3/2$ as we tune the value of $\Delta$ by tilting the sample. At low tilts, where the $N=1\uparrow$ character is favored, we encounter an incompressible state that we label ICP1. As $\Delta$ is decreased this state disappears into an isotropic compressible state that we label CP1. Then, for a {\it narrow but finite} range of $\Delta$, a second incompressible state appears and we label it ICP2. This incompressible state then transitions into a strongly anisotropic compressible state that we label ACP. Finally, at larger $\Delta$, this state transitions into a second compressible phase that we label CP2 (see Fig.~\ref{0-1}).

The first important consideration is to establish a correspondence between the locations of these different phases relative to the ideal point at which the single particle splitting, $\Delta$, vanishes. We associate the location of this point in experiment with the peak obtained from of the hysteretic sweeps of the magnetoresistance at high temperatures, depicted in Fig. \ref{Fig3}\textbf{E}, across $1 \leq \tilde{\nu} \leq 2$. The ICP2 phase occurs in the vicinity of such high temperature features, and therefore we locate the ICP2 phase to be in the vicinity of the single particle level crossing, as depicted in Fig.~\ref{0-1}.

We will now list a series of possible candidate phases and their properties and discuss which of these phases are consistent with those we are observing. At filling $\tilde{\nu}=3/2$, a large class of states can be considered by employing the composite fermion picture. We begin with phases that have a composite Fermi liquid (CFL) nature. Since we are considering a two-component system it is possible to have CFL phases with two Fermi surfaces (in analogy to the $N=0$ Landau level in the limit of small Zeeman splitting).\cite{du:1995,kukushkin:1999,tiemann:2012} We label these phases as CFL$(\nu_{h0},\nu_{h1})$ by the corresponding partial {\it hole fillings} of the $N=0/1$ components. These phases have a compressible nature, featuring a finite metallic-like resistivity $\rho_{xx}\neq 0$ and an unquantized Hall resistivity $\rho_{xy}$.

\begin{figure}[h]
\centering
\includegraphics{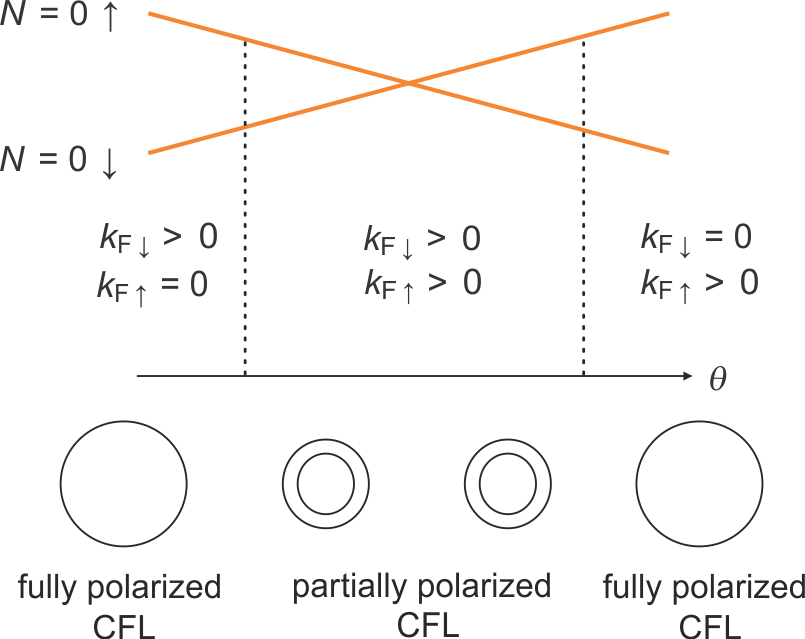}
\caption{Continuous depolarization in the auxiliary problem of a two component half-filled $N=0$ LL.}
\label{CFLs}
\end{figure}

These phases are natural candidates for the CP1 and CP2 phases we observe. The CFL$(1/2,0)$ is undoubtedly a good candidate description of the CP2 phase we are observing, since this phase appears in the limit of large splitting and when the chemical potential lies in the $N=0$LL. The phase CP1, on the other hard, is more difficult to understand. It nominally appears in the region in which the chemical potential lies in the $N=1\uparrow$ level but only within a range that lies in the vicinity of the level crossing. It is possible that this phase corresponds to a state with two Fermi surfaces, namely CFL$(\nu_{h0},\nu_{h1})$ with $\nu_{h0,h1}\neq0$. Such two component state have been considered theoretically in Refs.~\onlinecite{barkeshli:2016} and \onlinecite{zaletel:2018}. Moreover a study of bilayer graphene system that also realizes an analogous level crossing between $N=0$ and $N=1$ Landau levels, reported a gradual depolarization of the two components at half-filling and also considered the above states as partially polarized candidate phases~\cite{zibrov:2017}. We would like to offer an argument for why they might be energetically plausible in the present context. To do so, we consider an auxiliary problem in which we have a system of two levels with opposite spins but the same orbital character $N=0$. It is well established experimentally and theoretically that at $\tilde{\nu}=1/2$ such a system is described by an unpolarized Fermi sea in which both components have the same density in the limit of $\Delta=0$. It is also well established that the polarization, $p$, as a function of $\Delta$ in such a system is a continuous function of $\Delta$ that interpolates between the two fully polarized states at large $|\Delta|$ as depicted in Fig.~\ref{CFLs}. Now, the problem we are considering can be viewed as one that is perturbed away from this auxiliary problem by tuning the Haldane pseudpotentials for a single component from the values corresponding to $N=0$ into those of $N=1$. Such perturbation is by no means small, but one could imagine that the correlations that help to establish the continuous depolarization auxiliary problem could survive against such perturbation. More physically, we can say that there is a composite fermion kinetic energy loss associated with depolarizing the two component Fermi sea, that competes with other potential energy gains such as the correlation energy gained by polarizing into a single component and pairing into the Moore-Read state. The question of how such energy competition is settled is very non-trivial but could be addressed in future numerical studies.

Another class of closely related states are those obtained by pairing of the composite fermions in a CFL$({\tilde{\nu}_0,\tilde{\nu}_1})$ phase. These states will be incompressible in the sense that they will feature a vanishing $\rho_{xx}$ and a quantized $\rho_{xy}[h/e^2]=2/5 $ at low temperatures. This is true even if only one of the two components forms a paired state while the other remains in a Fermi surface state, such as in the recently proposed $Z_2$ exciton metal~\cite{barkeshli:2016,zaletel:2018}. This property can be economically understood by appealing to the parton construction of the composite fermion states (see e.g. Ref.~\onlinecite{barkeshli:2012}) in conjunction with the Ioffe-Larkin rule~\cite{ioffe:1989}, according to which we view the physical electron as: 

\be
c^\dagger_{N}=\psi_{N}^\dagger b^\dagger, \ N=\{0\downarrow,1\uparrow\}.
\ee

\noindent where $b^\dagger$ is a boson that carries physical charge $1$ and $\psi_{N}^\dagger$ are charge neutral composite fermions that carry the Landau level flavor degree of freedom $N$. The boson forms a $\nu=1/2$ Laughlin state, and the fermions can form states at effective zero magnetic field such as Fermi surfaces or paired states. According to the Ioffe-Larkin rule the physical resistivity is obtained by adding the resistivity tensors of the partons as follows: $\rho=\rho_b+\rho_\psi$. On the one hand the resistivity tensor of the bosonic sector is simply the one of the Laughlin state at $\nu=1/2$. The fermions will have a resistivity that would correspond to state a at effective zero magnetic field. If the fermions are paired the net resistivity will vanish. This is true even if only one component is paired and the other forms a Fermi surface,\cite{note1} such as the case of the $Z_2$ exciton metal proposed in Ref.~\onlinecite{barkeshli:2016}.

There is a plethora of states that one could obtain by considering different specific pairing channels of composite fermions. From the point of view of electric transport all of them will be candidates for the ICP1 and ICP2 phases. The most natural candidate state for the ICP1 phase is undoubtedly the Moore-Read state (or its particle-hole conjugate) which can be viewed as $p+ip$ weakly-paired states of the CFL$(0,1/2)$, since this is the leading candidate to explain the even denominator state of an isolated $N=1$ Landau level. Also, it is possible that near the boundary of the ICP1 phase and the CP1 phase the partially polarized $Z_2$ exciton metal proposed in Ref.~\onlinecite{barkeshli:2016} might be realized, since this state would display essentially similar features in charge transport as the Moore-Read state.

The ICP2 phase is much more non-trivial to understand, but we would like to elaborate on a few potential candidates. One possibility is again that this phase corresponds also to the Moore-Read state. Some studies have found that the Moore-Read state is enhanced near level crossings of sub-bands before suddenly disappearing.\cite{liu:2011} In light of this, an interesting possibility is that the Moore-Read state never truly disappears but its gap has a very non-monotonic behavior as a function of the level splitting so that the phase we call CP1 in an ideal limit is simply a weaker version of the MR state with a gap that has been washed out by disorder and temperature effects, which has a sudden revival near the level crossing. Another interesting possibility is that this state is a paired state that is sharply distinct from the Moore-Read state. Since this state occurs near the coincidence point of the two levels, a natural alternative candidate would be the analogue of the Halperin $331$ state which can also be understood as a paired state in the two component triplet $p+ip$ channel~\cite{read:2000} and has equal density for both components.

Finally, we discuss the strongly anisotropic state ACP that appears in between the incompressible state ICP2 and the compressible CP2 state. A natural candidate for a state with strong transport anisotropy in a half-filled Landau level is the stripe phase. This phase arises more naturally in higher Landau levels, but it is not un-common for it to be stabilized in the $N=1$ Landau level since it is understood to be in close energetic competition with the Moore-Read state.\cite{rezayi:2000} The feature that makes its appearance unexpected in the current case is that the phase seems favored near the level crossing with a $N=0$ Landau level. One interesting scenario is that this phase corresponds to a state with coherence between the $N=0\downarrow$ and $N=1\uparrow$ levels. As these orbitals carry different orbital angular momentum, a coherent superposition would break inversion and rotation symmetries, rendering the state with nematic and ferroelectric characteristics, explaining its anisotropic nature. In addition the opposite spin would endow the state with a finite magnetization in the plane orthogonal to the spin quantization axis dictated by the Zeeman energy. Proposals for related ferroelectric states in bilayer graphene have been made at integer filling factors (Ref. \onlinecite{cote:2010}) but there are to the best of our knowledge no studies addressing their energetic feasibility for half-filled Landau levels.

\bibliographystyle{apsrev}
\bibliography{mybibfile}

\begin{thebibliography}{40}
\expandafter\ifx\csname natexlab\endcsname\relax\def\natexlab#1{#1}\fi
\expandafter\ifx\csname bibnamefont\endcsname\relax
  \def\bibnamefont#1{#1}\fi
\expandafter\ifx\csname bibfnamefont\endcsname\relax
  \def\bibfnamefont#1{#1}\fi
\expandafter\ifx\csname citenamefont\endcsname\relax
  \def\citenamefont#1{#1}\fi
\expandafter\ifx\csname url\endcsname\relax
  \def\url#1{\texttt{#1}}\fi
\expandafter\ifx\csname urlprefix\endcsname\relax\def\urlprefix{URL }\fi
\providecommand{\bibinfo}[2]{#2}
\providecommand{\eprint}[2][]{\url{#2}}

\bibitem[{\citenamefont{Jain}(2007)}]{jain:2007}
\bibinfo{author}{\bibfnamefont{J.~K.} \bibnamefont{Jain}},
  \emph{\bibinfo{title}{Composite fermions}} (\bibinfo{publisher}{Cambridge
  University Press}, \bibinfo{year}{2007}).

\bibitem[{\citenamefont{Willett et~al.}(1987)\citenamefont{Willett, Eisenstein,
  St\"ormer, Tsui, Gossard, and English}}]{willett:1987}
\bibinfo{author}{\bibfnamefont{R.}~\bibnamefont{Willett}},
  \bibinfo{author}{\bibfnamefont{J.~P.} \bibnamefont{Eisenstein}},
  \bibinfo{author}{\bibfnamefont{H.~L.} \bibnamefont{St\"ormer}},
  \bibinfo{author}{\bibfnamefont{D.~C.} \bibnamefont{Tsui}},
  \bibinfo{author}{\bibfnamefont{A.~C.} \bibnamefont{Gossard}},
  \bibnamefont{and} \bibinfo{author}{\bibfnamefont{J.~H.}
  \bibnamefont{English}}, \bibinfo{journal}{Phys. Rev. Lett.}
  \textbf{\bibinfo{volume}{59}}, \bibinfo{pages}{1776} (\bibinfo{year}{1987}).

\bibitem[{\citenamefont{Ki et~al.}(2014)\citenamefont{Ki, Fal’ko, Abanin, and
  Morpurgo}}]{ki:2014}
\bibinfo{author}{\bibfnamefont{D.-K.} \bibnamefont{Ki}},
  \bibinfo{author}{\bibfnamefont{V.~I.} \bibnamefont{Fal’ko}},
  \bibinfo{author}{\bibfnamefont{D.~A.} \bibnamefont{Abanin}},
  \bibnamefont{and} \bibinfo{author}{\bibfnamefont{A.~F.}
  \bibnamefont{Morpurgo}}, \bibinfo{journal}{Nano Letters}
  \textbf{\bibinfo{volume}{14}}, \bibinfo{pages}{2135} (\bibinfo{year}{2014}).

\bibitem[{\citenamefont{Falson et~al.}(2015)\citenamefont{Falson, Maryenko,
  Friess, Zhang, Kozuka, Tsukazaki, Smet, and Kawasaki}}]{falson:2015a}
\bibinfo{author}{\bibfnamefont{J.}~\bibnamefont{Falson}},
  \bibinfo{author}{\bibfnamefont{D.}~\bibnamefont{Maryenko}},
  \bibinfo{author}{\bibfnamefont{B.}~\bibnamefont{Friess}},
  \bibinfo{author}{\bibfnamefont{D.}~\bibnamefont{Zhang}},
  \bibinfo{author}{\bibfnamefont{Y.}~\bibnamefont{Kozuka}},
  \bibinfo{author}{\bibfnamefont{A.}~\bibnamefont{Tsukazaki}},
  \bibinfo{author}{\bibfnamefont{J.~H.} \bibnamefont{Smet}}, \bibnamefont{and}
  \bibinfo{author}{\bibfnamefont{M.}~\bibnamefont{Kawasaki}},
  \bibinfo{journal}{Nature Physics} \textbf{\bibinfo{volume}{11}},
  \bibinfo{pages}{347} (\bibinfo{year}{2015}).

\bibitem[{\citenamefont{A.~Zibrov et~al.}(2017)\citenamefont{A.~Zibrov,
  Kometter, Zhou, M.~Spanton, Taniguchi, Watanabe, P.~Zaletel, and
  F.~Young}}]{zibrov:2017}
\bibinfo{author}{\bibfnamefont{A.}~\bibnamefont{A.~Zibrov}},
  \bibinfo{author}{\bibfnamefont{C.}~\bibnamefont{Kometter}},
  \bibinfo{author}{\bibfnamefont{H.}~\bibnamefont{Zhou}},
  \bibinfo{author}{\bibfnamefont{E.}~\bibnamefont{M.~Spanton}},
  \bibinfo{author}{\bibfnamefont{T.}~\bibnamefont{Taniguchi}},
  \bibinfo{author}{\bibfnamefont{K.}~\bibnamefont{Watanabe}},
  \bibinfo{author}{\bibfnamefont{M.}~\bibnamefont{P.~Zaletel}},
  \bibnamefont{and} \bibinfo{author}{\bibfnamefont{A.}~\bibnamefont{F.~Young}},
  \bibinfo{journal}{Nature} \textbf{\bibinfo{volume}{549}},
  \bibinfo{pages}{360} (\bibinfo{year}{2017}).

\bibitem[{\citenamefont{Li et~al.}(2017)\citenamefont{Li, Tan, Chen, Zeng,
  Taniguchi, Watanabe, Hone, and Dean}}]{li:2017}
\bibinfo{author}{\bibfnamefont{J.~I.~A.} \bibnamefont{Li}},
  \bibinfo{author}{\bibfnamefont{C.}~\bibnamefont{Tan}},
  \bibinfo{author}{\bibfnamefont{S.}~\bibnamefont{Chen}},
  \bibinfo{author}{\bibfnamefont{Y.}~\bibnamefont{Zeng}},
  \bibinfo{author}{\bibfnamefont{T.}~\bibnamefont{Taniguchi}},
  \bibinfo{author}{\bibfnamefont{K.}~\bibnamefont{Watanabe}},
  \bibinfo{author}{\bibfnamefont{J.}~\bibnamefont{Hone}}, \bibnamefont{and}
  \bibinfo{author}{\bibfnamefont{C.~R.} \bibnamefont{Dean}},
  \bibinfo{journal}{Science} \textbf{\bibinfo{volume}{358}},
  \bibinfo{pages}{648} (\bibinfo{year}{2017}).

\bibitem[{\citenamefont{Halperin et~al.}(1993)\citenamefont{Halperin, Lee, and
  Read}}]{halperin:1993}
\bibinfo{author}{\bibfnamefont{B.~I.} \bibnamefont{Halperin}},
  \bibinfo{author}{\bibfnamefont{P.~A.} \bibnamefont{Lee}}, \bibnamefont{and}
  \bibinfo{author}{\bibfnamefont{N.}~\bibnamefont{Read}},
  \bibinfo{journal}{Phys. Rev. B} \textbf{\bibinfo{volume}{47}},
  \bibinfo{pages}{7312} (\bibinfo{year}{1993}).

\bibitem[{\citenamefont{Read and Green}(2000)}]{read:2000}
\bibinfo{author}{\bibfnamefont{N.}~\bibnamefont{Read}} \bibnamefont{and}
  \bibinfo{author}{\bibfnamefont{D.}~\bibnamefont{Green}},
  \bibinfo{journal}{Phys. Rev. B} \textbf{\bibinfo{volume}{61}},
  \bibinfo{pages}{10267} (\bibinfo{year}{2000}).

\bibitem[{\citenamefont{Zhou et~al.}(2017)\citenamefont{Zhou, Kanoda, and
  Ng}}]{zhou:2017}
\bibinfo{author}{\bibfnamefont{Y.}~\bibnamefont{Zhou}},
  \bibinfo{author}{\bibfnamefont{K.}~\bibnamefont{Kanoda}}, \bibnamefont{and}
  \bibinfo{author}{\bibfnamefont{T.-K.} \bibnamefont{Ng}},
  \bibinfo{journal}{Rev. Mod. Phys.} \textbf{\bibinfo{volume}{89}},
  \bibinfo{pages}{025003} (\bibinfo{year}{2017}).

\bibitem[{\citenamefont{Alicea}(2012)}]{alicea:2012}
\bibinfo{author}{\bibfnamefont{J.}~\bibnamefont{Alicea}},
  \bibinfo{journal}{Reports on Progress in Physics}
  \textbf{\bibinfo{volume}{75}}, \bibinfo{pages}{076501}
  (\bibinfo{year}{2012}).

\bibitem[{\citenamefont{{Kasahara} et~al.}(2017)\citenamefont{{Kasahara},
  {Sugii}, {Ohnishi}, {Shimozawa}, {Yamashita}, {Kurita}, {Tanaka}, {Nasu},
  {Motome}, {Shibauchi} et~al.}}]{kasahara:2017}
\bibinfo{author}{\bibfnamefont{Y.}~\bibnamefont{{Kasahara}}},
  \bibinfo{author}{\bibfnamefont{K.}~\bibnamefont{{Sugii}}},
  \bibinfo{author}{\bibfnamefont{T.}~\bibnamefont{{Ohnishi}}},
  \bibinfo{author}{\bibfnamefont{M.}~\bibnamefont{{Shimozawa}}},
  \bibinfo{author}{\bibfnamefont{M.}~\bibnamefont{{Yamashita}}},
  \bibinfo{author}{\bibfnamefont{N.}~\bibnamefont{{Kurita}}},
  \bibinfo{author}{\bibfnamefont{H.}~\bibnamefont{{Tanaka}}},
  \bibinfo{author}{\bibfnamefont{J.}~\bibnamefont{{Nasu}}},
  \bibinfo{author}{\bibfnamefont{Y.}~\bibnamefont{{Motome}}},
  \bibinfo{author}{\bibfnamefont{T.}~\bibnamefont{{Shibauchi}}},
  \bibnamefont{et~al.}, \bibinfo{journal}{ArXiv e-prints}
  (\bibinfo{year}{2017}), \eprint{1709.10286}.

\bibitem[{\citenamefont{Nayak et~al.}(2008)\citenamefont{Nayak, Simon, Stern,
  Freedman, and Das~Sarma}}]{nayak:2008}
\bibinfo{author}{\bibfnamefont{C.}~\bibnamefont{Nayak}},
  \bibinfo{author}{\bibfnamefont{S.~H.} \bibnamefont{Simon}},
  \bibinfo{author}{\bibfnamefont{A.}~\bibnamefont{Stern}},
  \bibinfo{author}{\bibfnamefont{M.}~\bibnamefont{Freedman}}, \bibnamefont{and}
  \bibinfo{author}{\bibfnamefont{S.}~\bibnamefont{Das~Sarma}},
  \bibinfo{journal}{Rev. Mod. Phys.} \textbf{\bibinfo{volume}{80}},
  \bibinfo{pages}{1083} (\bibinfo{year}{2008}).

\bibitem[{\citenamefont{Barkeshli et~al.}(2016)\citenamefont{Barkeshli, Nayak,
  Papic, Young, and Zaletel}}]{barkeshli:2016}
\bibinfo{author}{\bibfnamefont{M.}~\bibnamefont{Barkeshli}},
  \bibinfo{author}{\bibfnamefont{C.}~\bibnamefont{Nayak}},
  \bibinfo{author}{\bibfnamefont{Z.}~\bibnamefont{Papic}},
  \bibinfo{author}{\bibfnamefont{A.}~\bibnamefont{Young}}, \bibnamefont{and}
  \bibinfo{author}{\bibfnamefont{M.}~\bibnamefont{Zaletel}},
  \bibinfo{journal}{arXiv preprint arXiv:1611.01171}  (\bibinfo{year}{2016}).

\bibitem[{\citenamefont{{Zaletel} et~al.}(2018)\citenamefont{{Zaletel},
  {Geraedts}, {Papi{\'c}}, and {Rezayi}}}]{zaletel:2018}
\bibinfo{author}{\bibfnamefont{M.~P.} \bibnamefont{{Zaletel}}},
  \bibinfo{author}{\bibfnamefont{S.}~\bibnamefont{{Geraedts}}},
  \bibinfo{author}{\bibfnamefont{Z.}~\bibnamefont{{Papi{\'c}}}},
  \bibnamefont{and} \bibinfo{author}{\bibfnamefont{E.~H.}
  \bibnamefont{{Rezayi}}}, \bibinfo{journal}{ArXiv e-prints}
  (\bibinfo{year}{2018}), \eprint{1803.08077}.

\bibitem[{\citenamefont{Hunt et~al.}(2017)\citenamefont{Hunt, Li, Zibrov, Wang,
  Taniguchi, Watanabe, Hone, Dean, Zaletel, Ashoori et~al.}}]{hunt:2017}
\bibinfo{author}{\bibfnamefont{B.}~\bibnamefont{Hunt}},
  \bibinfo{author}{\bibfnamefont{J.}~\bibnamefont{Li}},
  \bibinfo{author}{\bibfnamefont{A.}~\bibnamefont{Zibrov}},
  \bibinfo{author}{\bibfnamefont{L.}~\bibnamefont{Wang}},
  \bibinfo{author}{\bibfnamefont{T.}~\bibnamefont{Taniguchi}},
  \bibinfo{author}{\bibfnamefont{K.}~\bibnamefont{Watanabe}},
  \bibinfo{author}{\bibfnamefont{J.}~\bibnamefont{Hone}},
  \bibinfo{author}{\bibfnamefont{C.}~\bibnamefont{Dean}},
  \bibinfo{author}{\bibfnamefont{M.}~\bibnamefont{Zaletel}},
  \bibinfo{author}{\bibfnamefont{R.}~\bibnamefont{Ashoori}},
  \bibnamefont{et~al.}, \bibinfo{journal}{Nature communications}
  \textbf{\bibinfo{volume}{8}}, \bibinfo{pages}{948} (\bibinfo{year}{2017}).

\bibitem[{\citenamefont{Novoselov et~al.}(2006)\citenamefont{Novoselov, McCann,
  Morozov, Fal’ko, Katsnelson, Zeitler, Jiang, Schedin, and
  Geim}}]{novoselov:2006}
\bibinfo{author}{\bibfnamefont{K.~S.} \bibnamefont{Novoselov}},
  \bibinfo{author}{\bibfnamefont{E.}~\bibnamefont{McCann}},
  \bibinfo{author}{\bibfnamefont{S.}~\bibnamefont{Morozov}},
  \bibinfo{author}{\bibfnamefont{V.~I.} \bibnamefont{Fal’ko}},
  \bibinfo{author}{\bibfnamefont{M.}~\bibnamefont{Katsnelson}},
  \bibinfo{author}{\bibfnamefont{U.}~\bibnamefont{Zeitler}},
  \bibinfo{author}{\bibfnamefont{D.}~\bibnamefont{Jiang}},
  \bibinfo{author}{\bibfnamefont{F.}~\bibnamefont{Schedin}}, \bibnamefont{and}
  \bibinfo{author}{\bibfnamefont{A.}~\bibnamefont{Geim}},
  \bibinfo{journal}{Nature physics} \textbf{\bibinfo{volume}{2}},
  \bibinfo{pages}{177} (\bibinfo{year}{2006}).

\bibitem[{\citenamefont{McCann and Fal'ko}(2006)}]{mccann:2006}
\bibinfo{author}{\bibfnamefont{E.}~\bibnamefont{McCann}} \bibnamefont{and}
  \bibinfo{author}{\bibfnamefont{V.~I.} \bibnamefont{Fal'ko}},
  \bibinfo{journal}{Phys. Rev. Lett.} \textbf{\bibinfo{volume}{96}},
  \bibinfo{pages}{086805} (\bibinfo{year}{2006}).

\bibitem[{\citenamefont{C\^ot\'e et~al.}(2010)\citenamefont{C\^ot\'e, Lambert,
  Barlas, and MacDonald}}]{cote:2010}
\bibinfo{author}{\bibfnamefont{R.}~\bibnamefont{C\^ot\'e}},
  \bibinfo{author}{\bibfnamefont{J.}~\bibnamefont{Lambert}},
  \bibinfo{author}{\bibfnamefont{Y.}~\bibnamefont{Barlas}}, \bibnamefont{and}
  \bibinfo{author}{\bibfnamefont{A.~H.} \bibnamefont{MacDonald}},
  \bibinfo{journal}{Phys. Rev. B} \textbf{\bibinfo{volume}{82}},
  \bibinfo{pages}{035445} (\bibinfo{year}{2010}).

\bibitem[{\citenamefont{C\^ot\'e et~al.}(2011)\citenamefont{C\^ot\'e, Fouquet,
  and Luo}}]{cote:2011}
\bibinfo{author}{\bibfnamefont{R.}~\bibnamefont{C\^ot\'e}},
  \bibinfo{author}{\bibfnamefont{J.~P.} \bibnamefont{Fouquet}},
  \bibnamefont{and} \bibinfo{author}{\bibfnamefont{W.}~\bibnamefont{Luo}},
  \bibinfo{journal}{Phys. Rev. B} \textbf{\bibinfo{volume}{84}},
  \bibinfo{pages}{235301} (\bibinfo{year}{2011}).

\bibitem[{\citenamefont{De~Poortere et~al.}(2000)\citenamefont{De~Poortere,
  Tutuc, Papadakis, and Shayegan}}]{depoortere:2000}
\bibinfo{author}{\bibfnamefont{E.~P.} \bibnamefont{De~Poortere}},
  \bibinfo{author}{\bibfnamefont{E.}~\bibnamefont{Tutuc}},
  \bibinfo{author}{\bibfnamefont{S.~J.} \bibnamefont{Papadakis}},
  \bibnamefont{and} \bibinfo{author}{\bibfnamefont{M.}~\bibnamefont{Shayegan}},
  \bibinfo{journal}{Science} \textbf{\bibinfo{volume}{290}},
  \bibinfo{pages}{1546} (\bibinfo{year}{2000}).

\bibitem[{\citenamefont{Gokmen et~al.}(2010)\citenamefont{Gokmen, Padmanabhan,
  and Shayegan}}]{gokmen:2010}
\bibinfo{author}{\bibfnamefont{T.}~\bibnamefont{Gokmen}},
  \bibinfo{author}{\bibfnamefont{M.}~\bibnamefont{Padmanabhan}},
  \bibnamefont{and} \bibinfo{author}{\bibfnamefont{M.}~\bibnamefont{Shayegan}},
  \bibinfo{journal}{Phys. Rev. B} \textbf{\bibinfo{volume}{81}},
  \bibinfo{pages}{235305} (\bibinfo{year}{2010}).

\bibitem[{\citenamefont{Maryenko et~al.}(2014)\citenamefont{Maryenko, Falson,
  Kozuka, Tsukazaki, and Kawasaki}}]{maryenko:2014}
\bibinfo{author}{\bibfnamefont{D.}~\bibnamefont{Maryenko}},
  \bibinfo{author}{\bibfnamefont{J.}~\bibnamefont{Falson}},
  \bibinfo{author}{\bibfnamefont{Y.}~\bibnamefont{Kozuka}},
  \bibinfo{author}{\bibfnamefont{A.}~\bibnamefont{Tsukazaki}},
  \bibnamefont{and} \bibinfo{author}{\bibfnamefont{M.}~\bibnamefont{Kawasaki}},
  \bibinfo{journal}{Phys. Rev. B} \textbf{\bibinfo{volume}{90}},
  \bibinfo{pages}{245303} (\bibinfo{year}{2014}).

\bibitem[{\citenamefont{Giuliani and Vignale}(2005)}]{giuliani:2005}
\bibinfo{author}{\bibfnamefont{G.}~\bibnamefont{Giuliani}} \bibnamefont{and}
  \bibinfo{author}{\bibfnamefont{G.}~\bibnamefont{Vignale}},
  \emph{\bibinfo{title}{Quantum theory of the electron liquid}}
  (\bibinfo{publisher}{Cambridge university press}, \bibinfo{year}{2005}).

\bibitem[{\citenamefont{Moore and Read}(1991)}]{moore:1991}
\bibinfo{author}{\bibfnamefont{G.~W.} \bibnamefont{Moore}} \bibnamefont{and}
  \bibinfo{author}{\bibfnamefont{N.}~\bibnamefont{Read}},
  \bibinfo{journal}{Nucl. Phys.} \textbf{\bibinfo{volume}{B360}},
  \bibinfo{pages}{362} (\bibinfo{year}{1991}).

\bibitem[{\citenamefont{Kukushkin et~al.}(1999)\citenamefont{Kukushkin,
  v.~Klitzing, and Eberl}}]{kukushkin:1999}
\bibinfo{author}{\bibfnamefont{I.~V.} \bibnamefont{Kukushkin}},
  \bibinfo{author}{\bibfnamefont{K.}~\bibnamefont{v.~Klitzing}},
  \bibnamefont{and} \bibinfo{author}{\bibfnamefont{K.}~\bibnamefont{Eberl}},
  \bibinfo{journal}{Phys. Rev. Lett.} \textbf{\bibinfo{volume}{82}},
  \bibinfo{pages}{3665} (\bibinfo{year}{1999}).

\bibitem[{\citenamefont{Tiemann et~al.}(2012)\citenamefont{Tiemann, Gamez,
  Kumada, and Muraki}}]{tiemann:2012}
\bibinfo{author}{\bibfnamefont{L.}~\bibnamefont{Tiemann}},
  \bibinfo{author}{\bibfnamefont{G.}~\bibnamefont{Gamez}},
  \bibinfo{author}{\bibfnamefont{N.}~\bibnamefont{Kumada}}, \bibnamefont{and}
  \bibinfo{author}{\bibfnamefont{K.}~\bibnamefont{Muraki}},
  \bibinfo{journal}{Science} \textbf{\bibinfo{volume}{335}},
  \bibinfo{pages}{828} (\bibinfo{year}{2012}).

\bibitem[{\citenamefont{Liu et~al.}(2011)\citenamefont{Liu, Kamburov, Shayegan,
  Pfeiffer, West, and Baldwin}}]{liu:2011}
\bibinfo{author}{\bibfnamefont{Y.}~\bibnamefont{Liu}},
  \bibinfo{author}{\bibfnamefont{D.}~\bibnamefont{Kamburov}},
  \bibinfo{author}{\bibfnamefont{M.}~\bibnamefont{Shayegan}},
  \bibinfo{author}{\bibfnamefont{L.~N.} \bibnamefont{Pfeiffer}},
  \bibinfo{author}{\bibfnamefont{K.~W.} \bibnamefont{West}}, \bibnamefont{and}
  \bibinfo{author}{\bibfnamefont{K.~W.} \bibnamefont{Baldwin}},
  \bibinfo{journal}{Phys. Rev. Lett.} \textbf{\bibinfo{volume}{107}},
  \bibinfo{pages}{176805} (\bibinfo{year}{2011}).

\bibitem[{\citenamefont{Fogler and Koulakov}(1997)}]{fogler:1997}
\bibinfo{author}{\bibfnamefont{M.~M.} \bibnamefont{Fogler}} \bibnamefont{and}
  \bibinfo{author}{\bibfnamefont{A.~A.} \bibnamefont{Koulakov}},
  \bibinfo{journal}{Phys. Rev. B} \textbf{\bibinfo{volume}{55}},
  \bibinfo{pages}{9326} (\bibinfo{year}{1997}).

\bibitem[{\citenamefont{Lilly et~al.}(1999{\natexlab{a}})\citenamefont{Lilly,
  Cooper, Eisenstein, Pfeiffer, and West}}]{lilly:1999a}
\bibinfo{author}{\bibfnamefont{M.~P.} \bibnamefont{Lilly}},
  \bibinfo{author}{\bibfnamefont{K.~B.} \bibnamefont{Cooper}},
  \bibinfo{author}{\bibfnamefont{J.~P.} \bibnamefont{Eisenstein}},
  \bibinfo{author}{\bibfnamefont{L.~N.} \bibnamefont{Pfeiffer}},
  \bibnamefont{and} \bibinfo{author}{\bibfnamefont{K.~W.} \bibnamefont{West}},
  \bibinfo{journal}{Phys. Rev. Lett.} \textbf{\bibinfo{volume}{82}},
  \bibinfo{pages}{394} (\bibinfo{year}{1999}{\natexlab{a}}).

\bibitem[{\citenamefont{Lilly et~al.}(1999{\natexlab{b}})\citenamefont{Lilly,
  Cooper, Eisenstein, Pfeiffer, and West}}]{lilly:1999b}
\bibinfo{author}{\bibfnamefont{M.~P.} \bibnamefont{Lilly}},
  \bibinfo{author}{\bibfnamefont{K.~B.} \bibnamefont{Cooper}},
  \bibinfo{author}{\bibfnamefont{J.~P.} \bibnamefont{Eisenstein}},
  \bibinfo{author}{\bibfnamefont{L.~N.} \bibnamefont{Pfeiffer}},
  \bibnamefont{and} \bibinfo{author}{\bibfnamefont{K.~W.} \bibnamefont{West}},
  \bibinfo{journal}{Phys. Rev. Lett.} \textbf{\bibinfo{volume}{83}},
  \bibinfo{pages}{824} (\bibinfo{year}{1999}{\natexlab{b}}).

\bibitem[{\citenamefont{Pan et~al.}(1999)\citenamefont{Pan, Du, Stormer, Tsui,
  Pfeiffer, Baldwin, and West}}]{pan:1999}
\bibinfo{author}{\bibfnamefont{W.}~\bibnamefont{Pan}},
  \bibinfo{author}{\bibfnamefont{R.~R.} \bibnamefont{Du}},
  \bibinfo{author}{\bibfnamefont{H.~L.} \bibnamefont{Stormer}},
  \bibinfo{author}{\bibfnamefont{D.~C.} \bibnamefont{Tsui}},
  \bibinfo{author}{\bibfnamefont{L.~N.} \bibnamefont{Pfeiffer}},
  \bibinfo{author}{\bibfnamefont{K.~W.} \bibnamefont{Baldwin}},
  \bibnamefont{and} \bibinfo{author}{\bibfnamefont{K.~W.} \bibnamefont{West}},
  \bibinfo{journal}{Phys. Rev. Lett.} \textbf{\bibinfo{volume}{83}},
  \bibinfo{pages}{820} (\bibinfo{year}{1999}).

\bibitem[{\citenamefont{Rezayi and Haldane}(2000)}]{rezayi:2000}
\bibinfo{author}{\bibfnamefont{E.~H.} \bibnamefont{Rezayi}} \bibnamefont{and}
  \bibinfo{author}{\bibfnamefont{F.~D.~M.} \bibnamefont{Haldane}},
  \bibinfo{journal}{Phys. Rev. Lett.} \textbf{\bibinfo{volume}{84}},
  \bibinfo{pages}{4685} (\bibinfo{year}{2000}).

\bibitem[{\citenamefont{Samkharadze et~al.}(2016)\citenamefont{Samkharadze,
  Schreiber, Gardner, Manfra, Fradkin, and Csathy}}]{samkharadze:2016}
\bibinfo{author}{\bibfnamefont{N.}~\bibnamefont{Samkharadze}},
  \bibinfo{author}{\bibfnamefont{K.}~\bibnamefont{Schreiber}},
  \bibinfo{author}{\bibfnamefont{G.}~\bibnamefont{Gardner}},
  \bibinfo{author}{\bibfnamefont{M.}~\bibnamefont{Manfra}},
  \bibinfo{author}{\bibfnamefont{E.}~\bibnamefont{Fradkin}}, \bibnamefont{and}
  \bibinfo{author}{\bibfnamefont{G.}~\bibnamefont{Csathy}},
  \bibinfo{journal}{Nature Physics} \textbf{\bibinfo{volume}{12}},
  \bibinfo{pages}{191} (\bibinfo{year}{2016}).

\bibitem[{\citenamefont{Falson and Kawasaki}(2018)}]{falson:2018}
\bibinfo{author}{\bibfnamefont{J.}~\bibnamefont{Falson}} \bibnamefont{and}
  \bibinfo{author}{\bibfnamefont{M.}~\bibnamefont{Kawasaki}},
  \bibinfo{journal}{Reports on Progress in Physics}
  \textbf{\bibinfo{volume}{81}}, \bibinfo{pages}{056501}
  (\bibinfo{year}{2018}).

\bibitem[{\citenamefont{Du et~al.}(1995)\citenamefont{Du, Yeh, Stormer, Tsui,
  Pfeiffer, and West}}]{du:1995}
\bibinfo{author}{\bibfnamefont{R.~R.} \bibnamefont{Du}},
  \bibinfo{author}{\bibfnamefont{A.~S.} \bibnamefont{Yeh}},
  \bibinfo{author}{\bibfnamefont{H.~L.} \bibnamefont{Stormer}},
  \bibinfo{author}{\bibfnamefont{D.~C.} \bibnamefont{Tsui}},
  \bibinfo{author}{\bibfnamefont{L.~N.} \bibnamefont{Pfeiffer}},
  \bibnamefont{and} \bibinfo{author}{\bibfnamefont{K.~W.} \bibnamefont{West}},
  \bibinfo{journal}{Phys. Rev. Lett.} \textbf{\bibinfo{volume}{75}},
  \bibinfo{pages}{3926} (\bibinfo{year}{1995}).

\bibitem[{\citenamefont{Barkeshli and McGreevy}(2012)}]{barkeshli:2012}
\bibinfo{author}{\bibfnamefont{M.}~\bibnamefont{Barkeshli}} \bibnamefont{and}
  \bibinfo{author}{\bibfnamefont{J.}~\bibnamefont{McGreevy}},
  \bibinfo{journal}{Phys. Rev. B} \textbf{\bibinfo{volume}{86}},
  \bibinfo{pages}{075136} (\bibinfo{year}{2012}).

\bibitem[{\citenamefont{Ioffe and Larkin}(1989)}]{ioffe:1989}
\bibinfo{author}{\bibfnamefont{L.~B.} \bibnamefont{Ioffe}} \bibnamefont{and}
  \bibinfo{author}{\bibfnamefont{A.~I.} \bibnamefont{Larkin}},
  \bibinfo{journal}{Phys. Rev. B} \textbf{\bibinfo{volume}{39}},
  \bibinfo{pages}{8988} (\bibinfo{year}{1989}).

\bibitem[{not()}]{note1}
\bibinfo{note}{Such case would be analogous to computing the net resistivity of
  a bilayer system in which one layer is in a superconducting state and the
  other in an ordinary metal. The current will be shunted by the
  superconducting component and hence the net resistivity vanishes.}

\bibitem[{\citenamefont{Luo and Chakraborty}(2016)}]{luo:2016a}
\bibinfo{author}{\bibfnamefont{W.}~\bibnamefont{Luo}} \bibnamefont{and}
  \bibinfo{author}{\bibfnamefont{T.}~\bibnamefont{Chakraborty}},
  \bibinfo{journal}{Phys. Rev. B} \textbf{\bibinfo{volume}{93}},
  \bibinfo{pages}{161103} (\bibinfo{year}{2016}).

\bibitem[{\citenamefont{Luo and Chakraborty}(2017)}]{luo:2017}
\bibinfo{author}{\bibfnamefont{W.}~\bibnamefont{Luo}} \bibnamefont{and}
  \bibinfo{author}{\bibfnamefont{T.}~\bibnamefont{Chakraborty}},
  \bibinfo{journal}{Phys. Rev. B} \textbf{\bibinfo{volume}{96}},
  \bibinfo{pages}{081108} (\bibinfo{year}{2017}).

\end{thebibliography}

\onecolumngrid

\section*{Extended data}\label{SampleParameters}

\begin{figure}[h]
\centering
\includegraphics{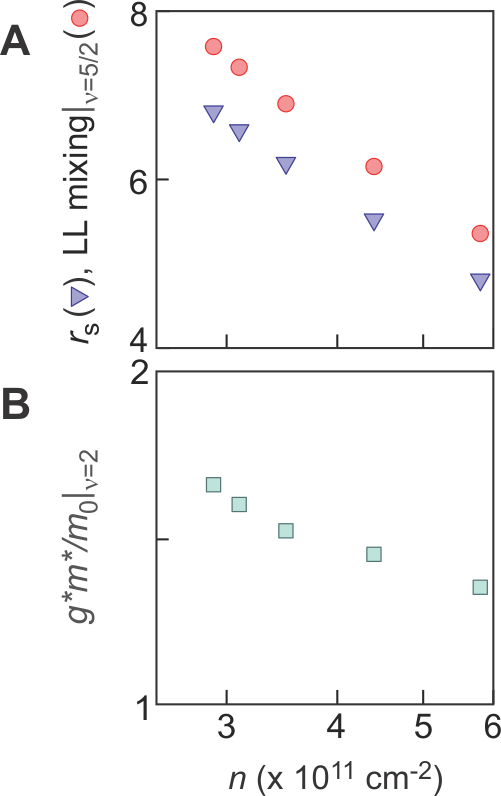}
\caption{Parameters of samples as a function of charge carrier density. (\textbf{A}) \textit{r}$_s$ (\Rs) and LL mixing at $\nu$~=~5/2 (\ECoul/\ECyc~=~16.6/$\sqrt{B}|_{\nu=5/2}$). The effect of LL mixing on the stability of FQH features has been discussed in Refs.~\onlinecite{luo:2016a} and~\onlinecite{luo:2017}. (\textbf{B}) \gm. We note the bulk parameters are \textit{m}$^*$~=~0.3$m_0$, \textit{g} $\approx$ 2 and dielectric constant $\epsilon = 8.5\epsilon_0$.}
\label{sampleparametersfig}
\end{figure}

\begin{figure}[h]
\centering
\includegraphics{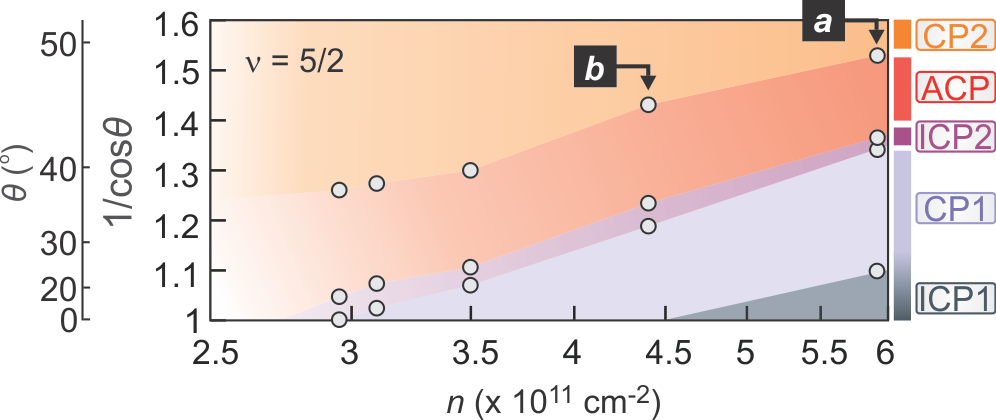}
\caption{Summary of the transitions at $\nu$~=~5/2 as a function of \textit{n} for multiple samples for 1/cos$\theta$.}
\label{angledependencephasediagram}
\end{figure}

\begin{figure}[h]
\centering
\includegraphics{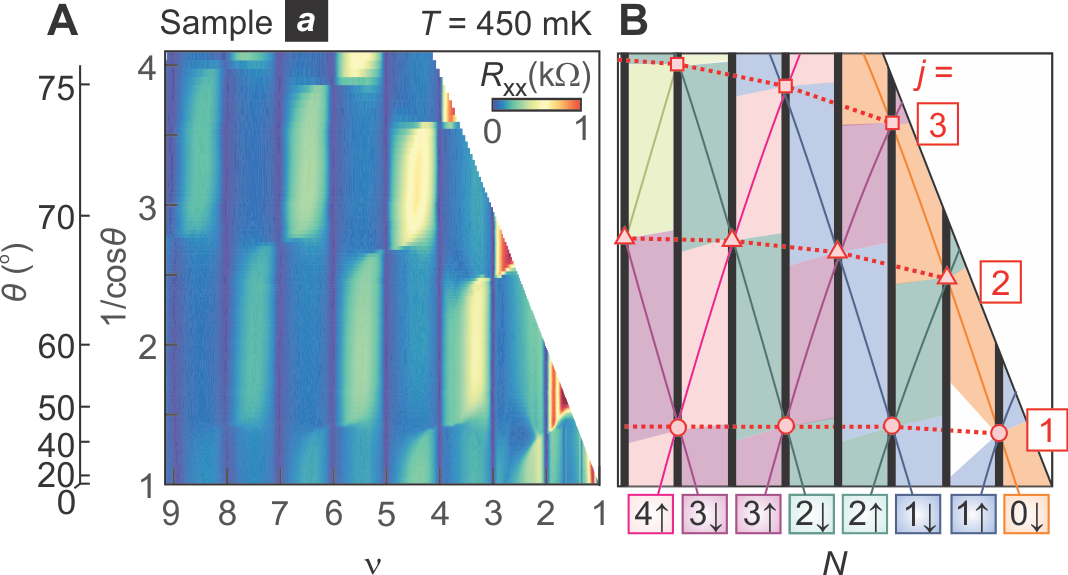}
\caption{Extended data set from Fig. \ref{Fig1}. (\textbf{A}) \textit{T}~=~450 mK mapping of the magnetotransport of sample \textit{a} (\textit{n}~=~5.8 \density) with (\textbf{B}) displaying the corresponding orbital quantum number and spin projection of the partially filled level. The first (\textit{j}~=~1, circles), second (\textit{j}~=~2, triangles) and third (\textit{j}~=~3, squares) coincidence positions are interpolated by dotted lines. This representation highlights that the \textit{j} coincidence positions slip to lower $\theta$ as $\nu$ is made small, which is attributed\cite{maryenko:2014} to a polarization dependent contribution to \gm.}
\label{450mkmapping}
\end{figure}

\begin{figure}[h]
\centering
\includegraphics{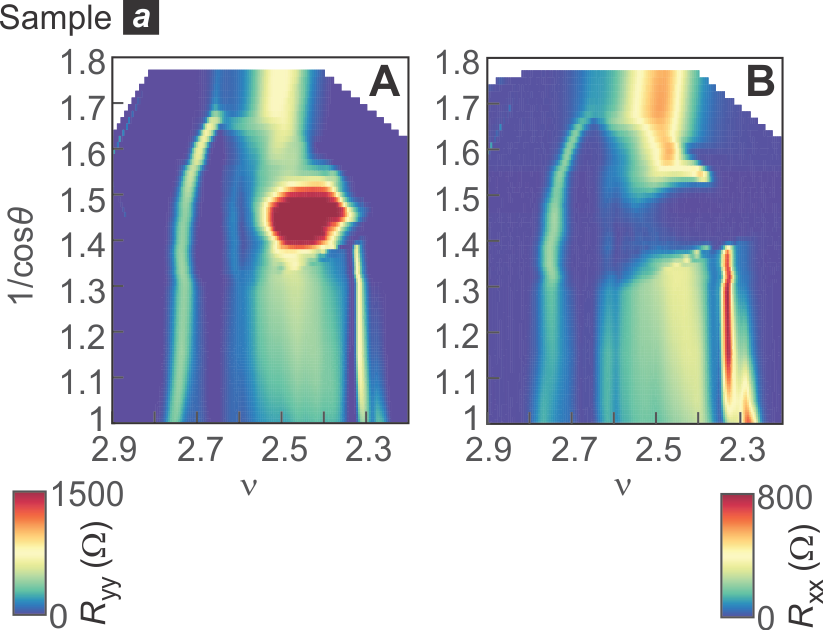}
\caption{Map of (\textbf{A}) \Rxx~ and (\textbf{B}) \Ryy~around $\nu$~=~5/2 for sample \textit{b} at \textit{T}~$\approx$~30 mK up to high field (\textit{B}~=~18~T). The ACP phase is completely resolved, along with the CP2 phase at higher $\theta$. Data shown in the main manuscript were taken in a \textit{B}~=~15~T, \textit{T}~$<$~20~mK cryostat.}
\label{samplebchip1}
\end{figure}

\begin{figure}[h]
\centering
\includegraphics{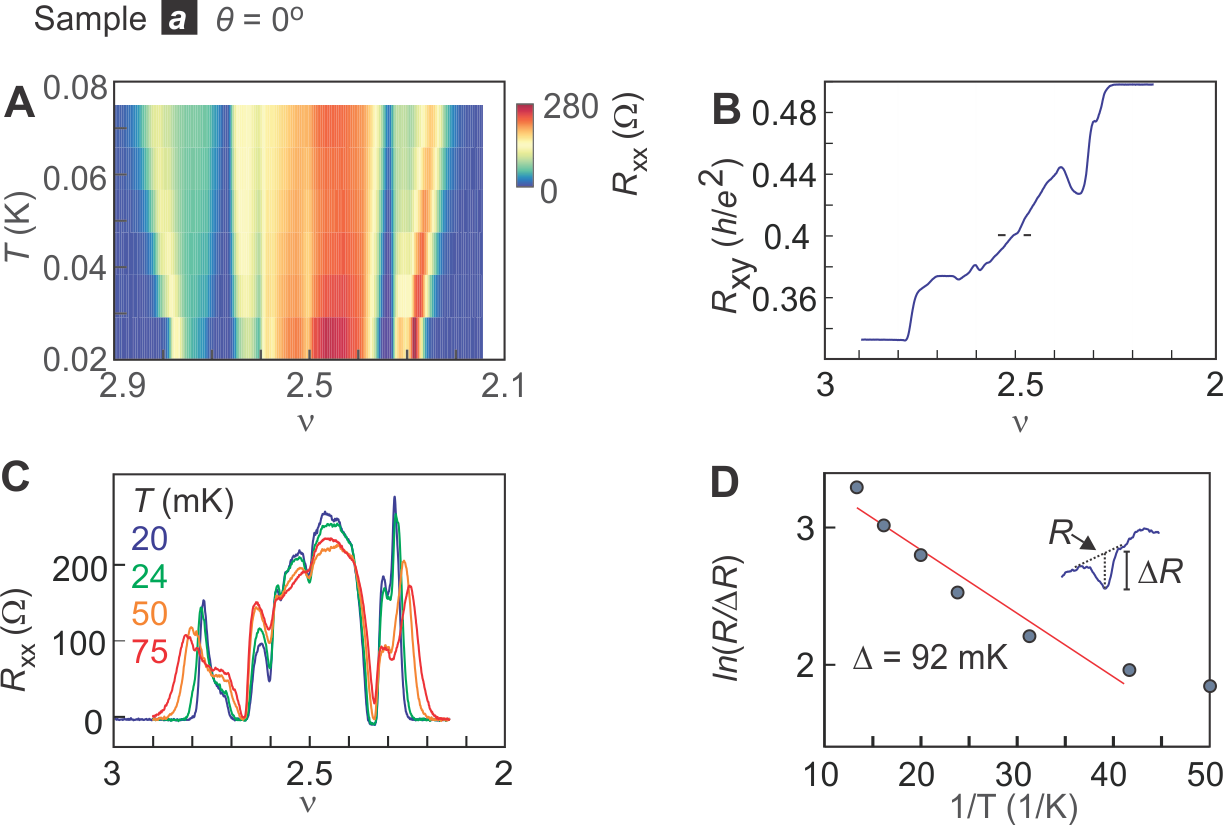}
\caption{(\textbf{A}) Magnetotransport as a function of \textit{T} when $\theta$~=~0\degree for sample \textit{a}. (\textbf{B}) \Rxy~at base temperature. (\textbf{C}) Individual line traces of the temperature dependent \Rxx~data. (\textbf{D}) Arrhenius plot of \textit{R}/$\Delta$\textit{R} of the $\nu$~=~5/2 resistance. The activation energy is estimated through \textit{R}/$\Delta$\textit{R} $\propto$ exp(-$\Delta_{\nu=5/2}/2T)$.}
\label{fivehalftempdependence}
\end{figure}

\begin{figure}[h]
\centering
\includegraphics{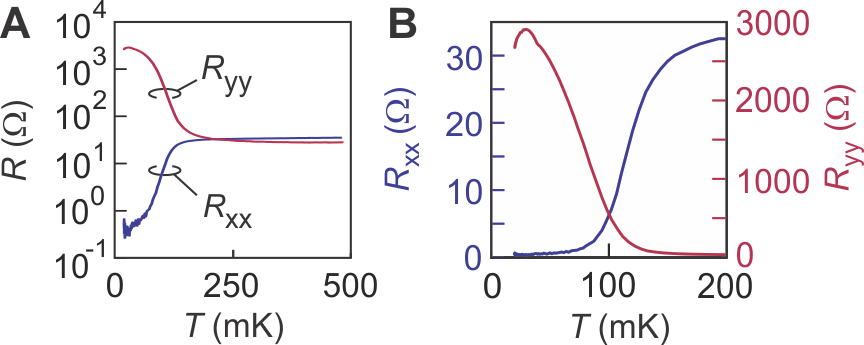}
\caption{Temperature dependence of the anisotropic phase for 1/cos$\theta$~=~1.33 and $\nu$~=~2.4 for two orthogonal crystal directions on (\textbf{A}) log and (\textbf{B}) linear scale.}
\label{tempdependencestripe}
\end{figure}

\begin{figure}[h]
\centering
\includegraphics{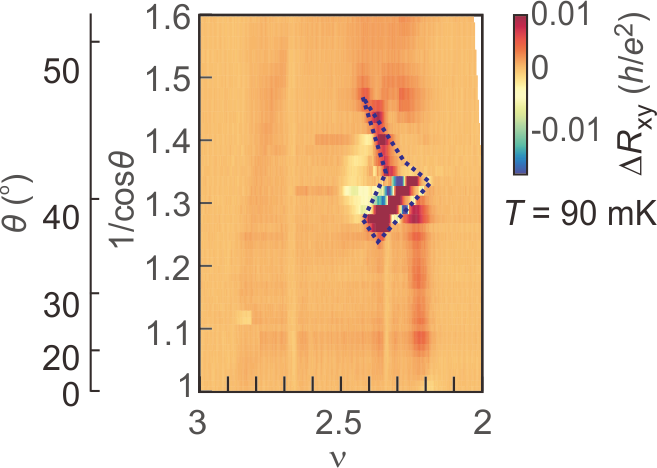}
\caption{Map of the hysteresis in the data presented in Fig. \ref{Fig3}\textbf{D} at \textit{T}~=~90~mK in \Rxy. The dotted blue region frames the hysteretic region that is incorporated into Fig. \ref{Fig4}\textbf{A}.}
\label{90mkhysteresis}
\end{figure}

\begin{figure}[h]
\centering
\includegraphics{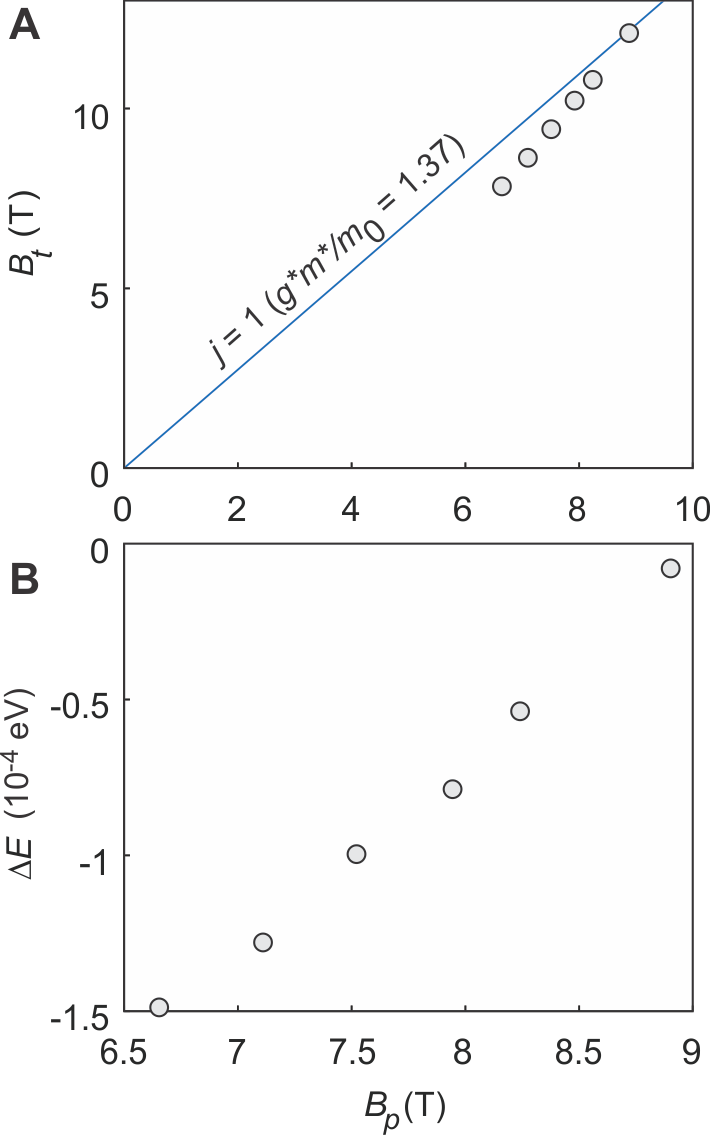}
\caption{(\textbf{A}) (Line) slope of the \textit{j}=1 coincidence assuming a constant \gm. (Open circles) Experimentally observed maxima in hysteresis of the resistance in Fig. \ref{Fig3}\textbf{E} in the \Bp-\Bt~plane taken on sample \textit{b}. (\textbf{B}) The deviation between these provides an estimate of the exchange energy corrections to the single particle energy levels.}
\label{exchangedata}
\end{figure}

\end{document}